\documentclass[preprint,11pt,3p]{elsarticle}

\usepackage{graphicx}
\usepackage{bm}

\usepackage{xspace} 
 
\newcommand{\MDRL}{MDRL\xspace} 

\makeatletter
\def\ps@pprintTitle{%
   \let\@oddhead\@empty
   \let\@evenhead\@empty
   \def\@oddfoot{\reset@font\hfil\thepage\hfil}
   \let\@evenfoot\@oddfoot
}
\makeatother

\usepackage{amsmath}
\DeclareMathOperator*{\argmax}{arg\,max}

\usepackage{booktabs}
\usepackage{url}

\usepackage[pagebackref=true]{hyperref}

\makeatletter
\newcommand\footnoteref[1]{\protected@xdef\@thefnmark{\ref{#1}}\@footnotemark}
\makeatother

\DeclareMathOperator{\E}{\mathbb{E}}

\usepackage{color}
\usepackage{amsmath}
\usepackage{amssymb}
\usepackage{amstext}
\usepackage{subcaption}
\usepackage[linesnumbered,ruled,vlined]{algorithm2e}

\newcommand\revise[1]{{\color{black}#1}} %

\begin{document}

\begin{frontmatter}

\title{A Survey and Critique of Multiagent Deep Reinforcement Learning\tnoteref{t1}}
\tnotetext[t1]{Earlier versions of this work had the title: ``Is multiagent deep reinforcement learning the answer or the question? A brief survey''}

\author{Pablo Hernandez-Leal, Bilal Kartal and Matthew E. Taylor\\
\texttt{\{pablo.hernandez,bilal.kartal,matthew.taylor\}@borealisai.com}
}

\address{
Borealis AI \\ 
Edmonton, Canada\\
}

\begin{abstract}
Deep reinforcement learning (RL) has achieved outstanding results in recent years. This has led to a dramatic increase in the number of applications and methods. Recent works have explored learning beyond single-agent scenarios and have considered multiagent learning (MAL) scenarios. Initial results report successes in complex multiagent domains, although there are several challenges to be addressed. The primary goal of this article is to provide a clear overview of current multiagent deep reinforcement learning (\MDRL) literature. Additionally, we complement the overview with a broader analysis: (i) we revisit previous key components, originally presented in MAL and RL, and highlight how they have been adapted to multiagent deep reinforcement learning settings. (ii) We provide general guidelines to new practitioners in the area: describing lessons learned from MDRL works, pointing to recent benchmarks, and outlining open avenues of research. (iii) We take a more critical tone raising practical challenges of MDRL (e.g., implementation and computational demands). We expect this article will help unify and motivate future research to take advantage of the abundant literature that exists (e.g., RL and MAL) in a joint effort to promote fruitful research in the multiagent community.
\end{abstract}

\end{frontmatter}

\section{Introduction}

Almost 20 years ago Stone and Veloso's seminal survey~\cite{Stone:2000ww} laid the groundwork for defining the area of multiagent systems (MAS) and its open problems in the context of AI. About ten years ago, Shoham, Powers, and Grenager~\cite{Shoham:2007vw} noted that the literature on multiagent learning (MAL) was growing and it was not possible to enumerate all relevant articles. Since then, the number of published MAL works continues to steadily rise, which led to different surveys on the area, ranging from analyzing the basics of MAL and their challenges~\cite{ALONSO:2002ko,Tuyls:2012up,Busoniu:2008bo}, to addressing specific subareas: game theory and MAL~\cite{Shoham:2007vw,nowe2012game}, cooperative scenarios~\cite{Panait:2005wj,Matignon:2012bj}, and evolutionary dynamics of MAL~\cite{Bloembergen:2015ei}. In just the last couple of years, three surveys related to MAL have been published: learning in non-stationary environments~\cite{HernandezLeal:2017wd}, agents modeling agents~\cite{Albrecht:2018dp}, and   transfer learning in multiagent RL~\cite{da2019survey}.

The research interest in MAL has been accompanied by successes in artificial intelligence, first, in single-agent video games~\cite{Mnih:2015jp}; more recently, in two-player games, for example, playing Go~\cite{Silver:2016hl,silver2017mastering}, poker~\cite{Moravcik:2017jb,brown2018superhuman}, and games of two competing teams, e.g., DOTA 2~\cite{openfive} and StarCraft II~\cite{alphastarblog}.

While different techniques and algorithms were used in the above scenarios, in general, they are all a combination of techniques from two main areas: reinforcement learning (RL)~\cite{sutton2018reinforcement} and deep learning~\cite{lecun2015deep,schmidhuber2015deep}. 

RL is an area of machine learning where an agent learns by interacting (i.e., taking actions) within a  dynamic environment. However, one of the main challenges to RL, and traditional machine learning in general, is the need for manually designing quality features on which to learn. Deep learning enables efficient representation learning, thus allowing the automatic discovery of features~\cite{lecun2015deep,schmidhuber2015deep}.
In recent years, deep learning has had successes in different areas such as computer vision and natural language processing~\cite{lecun2015deep,schmidhuber2015deep}.
One of the key aspects of deep learning is the use of \emph{neural networks} (NNs) that can find compact representations in high-dimensional data~\cite{Arulkumaran:2017jn}.

In deep reinforcement learning (DRL)~\cite{Arulkumaran:2017jn,franccois2018introduction} deep neural networks are trained to approximate the optimal policy and/or the value function. In this way the deep NN, serving as function approximator, enables powerful generalization.
One of the key advantages of DRL is that it enables RL to scale to problems with high-dimensional state and action spaces. However, most existing successful DRL applications so far have been on visual domains (e.g., Atari games), and there is still a lot of work to be done for more realistic applications~\cite{Yang:2018tu,zhao2018deep} with complex dynamics, which are not necessarily vision-based.

DRL has been regarded as an important component in constructing general AI systems~\cite{Lake:wm} and has been successfully integrated with other techniques, e.g., search~\cite{Silver:2016hl}, planning~\cite{Tamar:2016tz}, and more recently with multiagent systems, with an emerging area of \emph{multiagent deep reinforcement learning} \emph{(\MDRL)}\cite{papoudakis2019dealing,nguyen2018deep}.\footnote{We have noted inconsistency in abbreviations such as: D-MARL, MADRL, deep-multiagent RL and MA-DRL.}

Learning in multiagent settings is fundamentally more difficult than the single-agent case due to the presence of multiagent pathologies, e.g., the moving target problem (non-stationarity)~\cite{Shoham:2007vw,Busoniu:2008bo,HernandezLeal:2017wd}, curse of dimensionality~\cite{Shoham:2007vw,Busoniu:2008bo}, multiagent credit assignment~\cite{wolpert2002optimal,Agogino:2004ug}, global exploration~\cite{Matignon:2012bj}, and relative overgeneralization~\cite{Fulda:2006tu,Wei:2016wq,Palmer:2018wv}. Despite this complexity, top AI conferences like AAAI, ICML, ICLR, IJCAI and NeurIPS, and specialized conferences such as AAMAS, have published works reporting successes in \MDRL. In light of these works, we believe it is pertinent to first, have an overview of the recent \MDRL works, and second, understand how these recent works relate to the existing literature.

This article contributes to the state of the art with a brief survey of the current works in \MDRL in an effort to complement existing surveys on multiagent learning~\cite{Busoniu:2010ft,HernandezLeal:2017wd}, cooperative learning~\cite{Panait:2005wj,Matignon:2012bj}, agents modeling agents~\cite{Albrecht:2018dp}, knowledge reuse in multiagent RL~\cite{da2019survey}, and (single-agent) deep reinforcement learning~\cite{Arulkumaran:2017jn,li2017deep}.

\begin{figure}
\centering
\begin{subfigure}[]{.49\linewidth}
\centering
\includegraphics[scale=0.48]{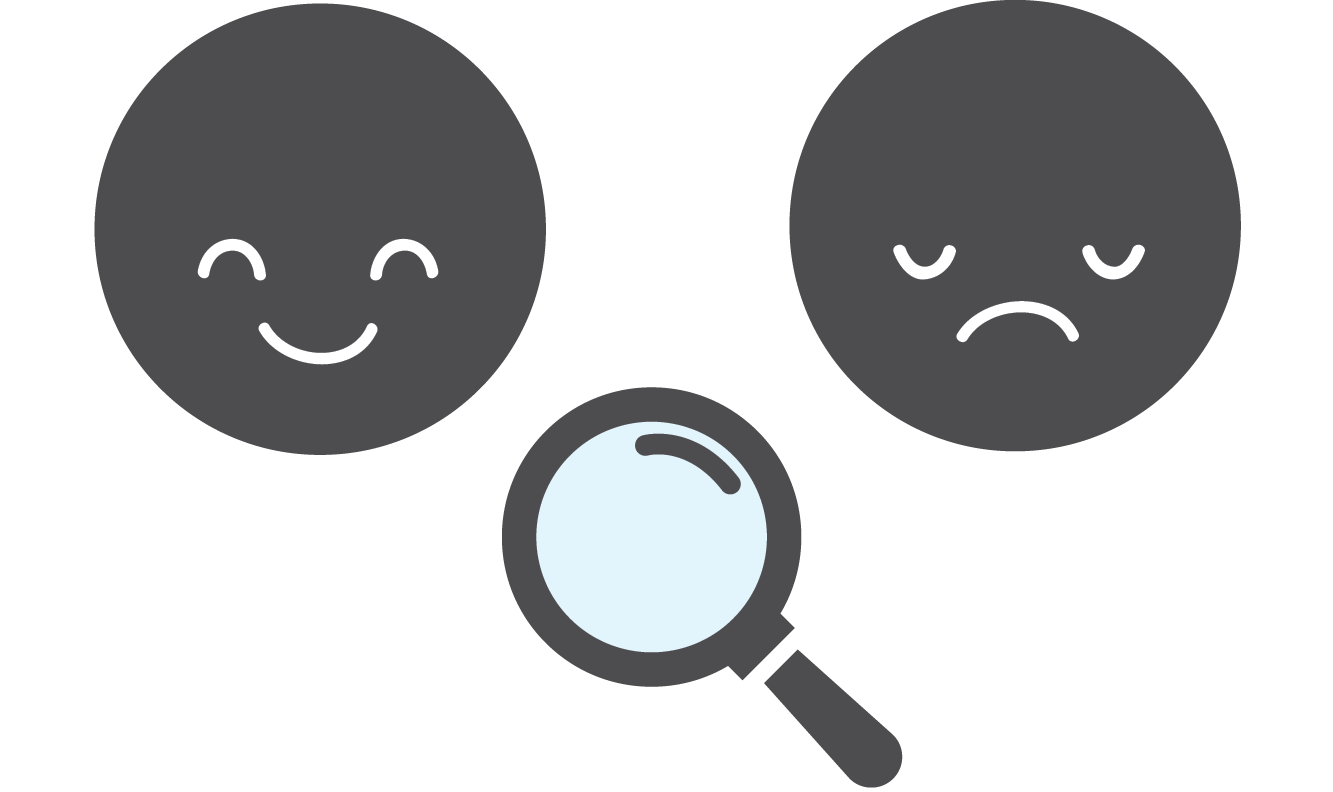}
\caption{Analysis of emergent behaviors}
\end{subfigure}
\begin{subfigure}[]{.49\linewidth}
\centering
\includegraphics[scale=0.48]{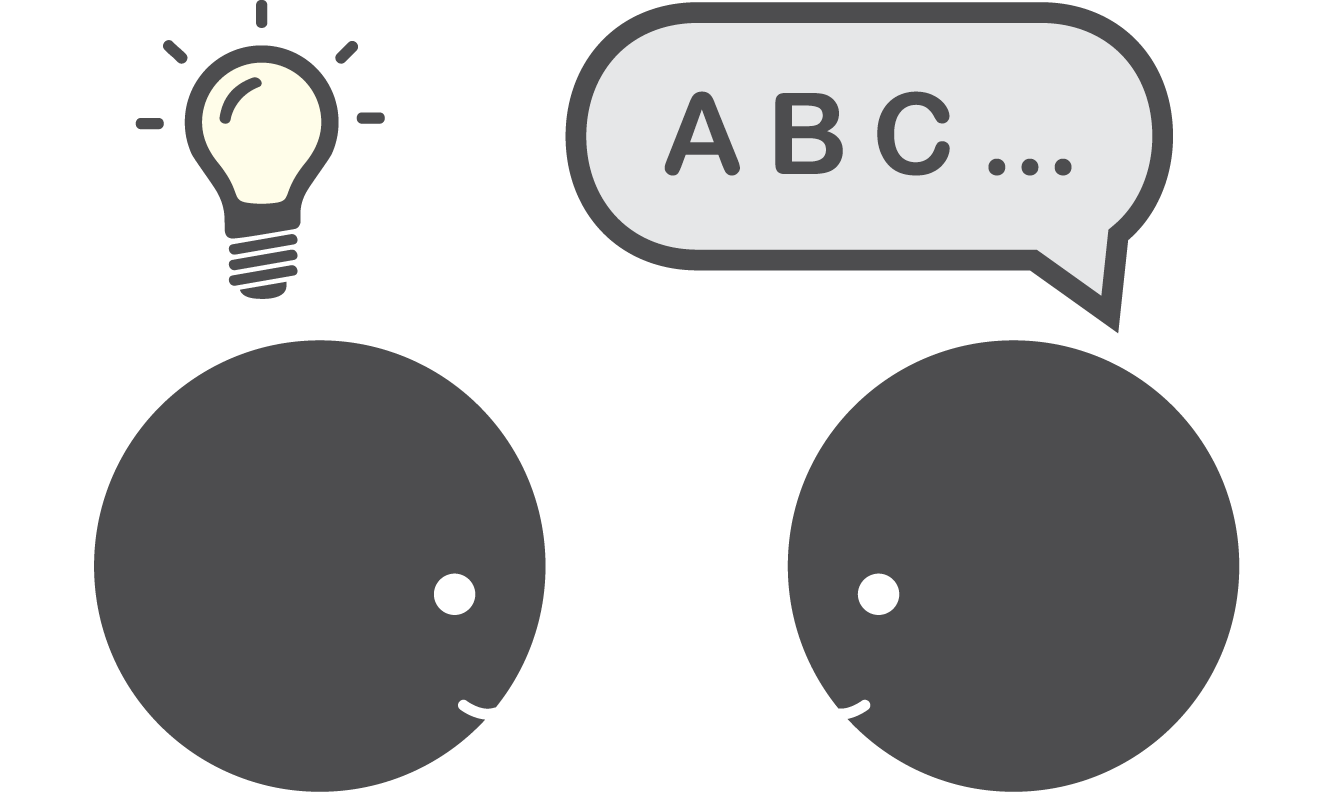}
\caption{Learning communication}
\end{subfigure} 
\begin{subfigure}[]{.49\linewidth}
\centering
\includegraphics[scale=0.48]{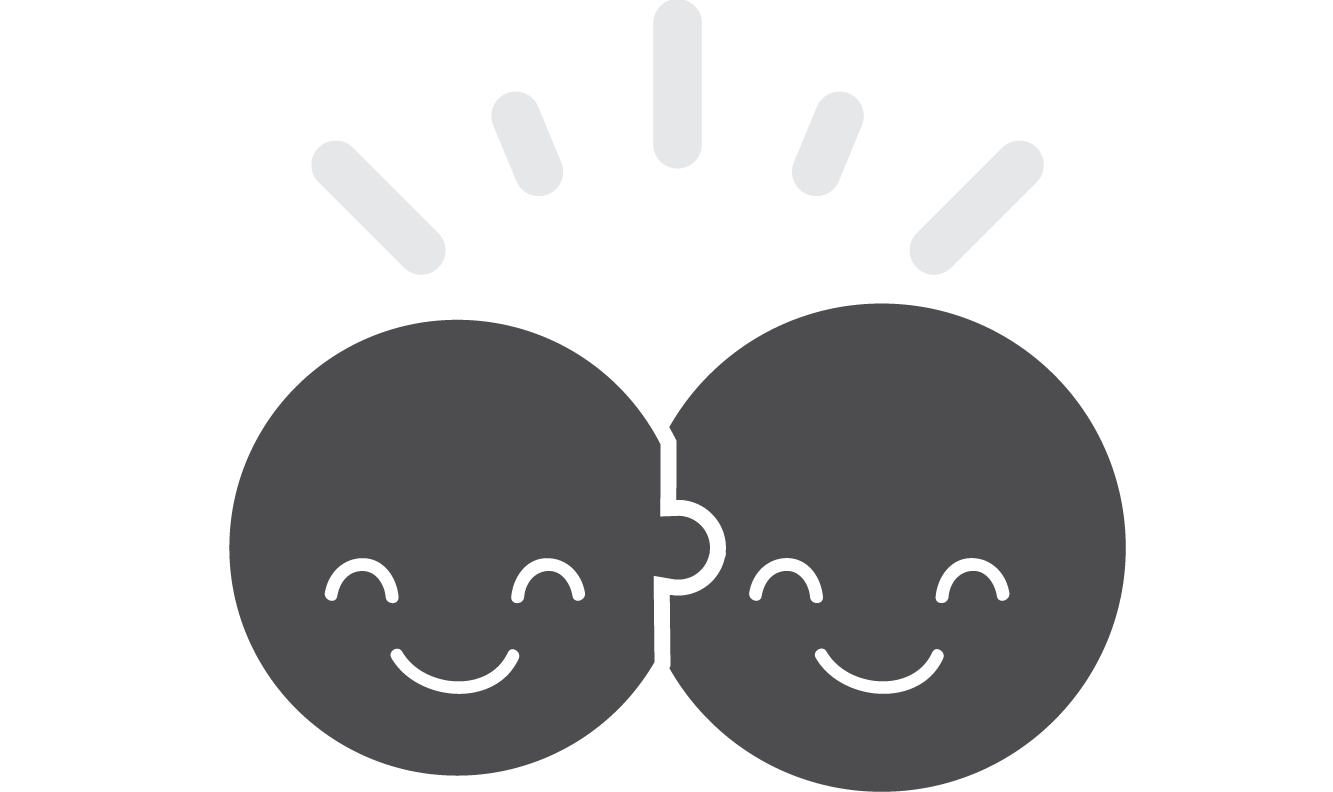}
\caption{Learning cooperation}
\end{subfigure}
\begin{subfigure}[]{.49\linewidth}
\centering
\includegraphics[scale=0.48]{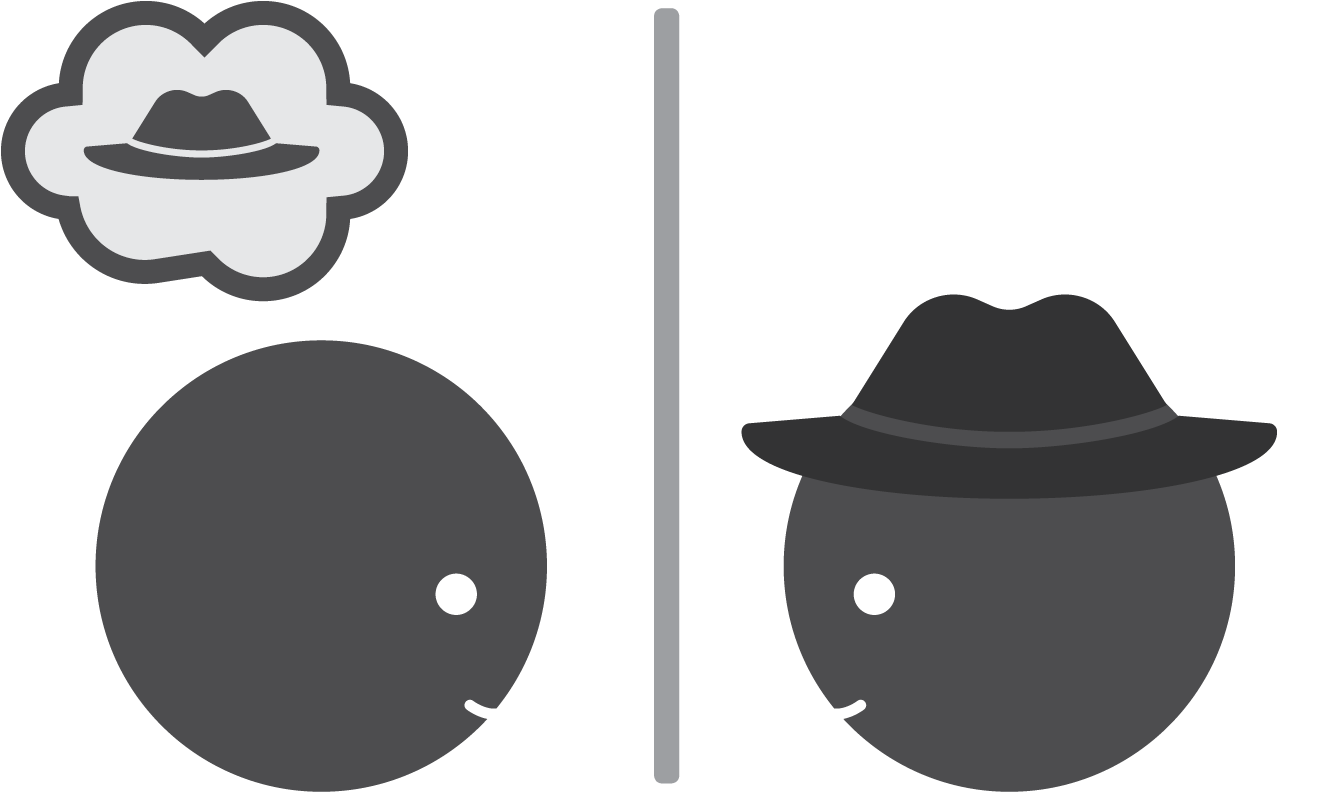}
\caption{Agents modeling agents}
\end{subfigure}
\caption{Categories of different \MDRL works. (a) Analysis of emergent behaviors: evaluate single-agent DRL algorithms in multiagent scenarios. (b) Learning communication: agents learn with actions and through messages. (c) Learning cooperation: agents learn to cooperate using only actions and (local) observations. (d) Agents modeling agents: agents reason about others to fulfill a task (e.g., cooperative or competitive). For a more detailed description see Sections~\ref{sec:emergentBehaviors}--\ref{sec:agentsModeling} and Tables~\ref{tab:emergent}--\ref{tab:modeling}.}
\label{fig:categories}
\end{figure}

First, we provide a short review of key algorithms in RL such as Q-learning and REINFORCE (see Section~\ref{sec:rl}). Second, we review DRL highlighting the challenges in this setting and reviewing recent works (see Section~\ref{sec:deeprl}). Third, we present the multiagent setting and give an overview of key challenges and results (see Section~\ref{sec:multiagentlearning}). Then, we present the identified four categories to group recent \MDRL works (see Figure~\ref{fig:categories}):
\begin{itemize} 

\item Analysis of emergent behaviors: evaluate single-agent DRL algorithms in multiagent scenarios (e.g., Atari games, social dilemmas, 3D competitive games).
\item Learning communication: agents learn communication protocols to solve cooperative tasks.
\item Learning cooperation: agents learn to cooperate using only actions and (local) observations.
\item Agents modeling agents: agents reason about others to fulfill a task (e.g., best response learners).

\end{itemize}

For each category we provide a description as well as outline the recent works (see Section~\ref{sec:categorization} and Tables~\ref{tab:emergent}--\ref{tab:modeling}). Then, we take a step back and reflect on how these new works relate to the existing literature. In that context, first, we present examples on how methods and algorithms originally introduced in RL and MAL were successfully been scaled to \MDRL (see Section~\ref{sec:bridging:examples}). Second, we provide some pointers for new practitioners in the area by describing general \emph{lessons learned} from the existing \MDRL works (see Section~\ref{sec:bridging:lessons}) and point to recent multiagent benchmarks (see Section~\ref{sec:benchmarksmdrl}). Third, we take a more critical view and describe practical challenges in MDRL, such as reproducibility, hyperparameter tunning, and computational demands (see Section~\ref{sec:challengesmdrl}). Then, we outline some open research questions (see Section~\ref{sec:bridging:questions}). Lastly, we present our conclusions from this work (see Section~\ref{sec:conclusions}).

Our goal is to outline a recent and active area (i.e., \MDRL), as well as to motivate future research to take advantage of the ample and existing literature in multiagent learning. We aim to enable researchers with experience in either DRL or MAL to gain a common understanding about recent works, and open problems in \MDRL, and to avoid having scattered sub-communities with little interaction~\cite{Shoham:2007vw,HernandezLeal:2017wd,Albrecht:2018dp,Darwiche:2018}.

\section{Single-agent learning}
\label{sec:background}

This section presents the formalism of reinforcement learning and its main components before outlining \emph{deep} reinforcement learning along with its particular challenges and recent algorithms. For a more detailed description we refer the reader to excellent books and surveys on the area~\cite{wiering2012reinforcement,sutton2018reinforcement,Arulkumaran:2017jn,kaelbling1996reinforcement,franccois2018introduction}.

\subsection{Reinforcement learning}
\label{sec:rl}

RL formalizes the interaction of an agent with an environment using a Markov decision process (MDP)~\cite{puterman1994markov}. An MDP is defined by the tuple $\langle \mathcal{S},\mathcal{A},R,T,\gamma \rangle$ where $\mathcal{S}$ represents a finite set of states. $\mathcal{A}$ represents a finite set of actions. The transition function $T : \mathcal{S} \times \mathcal{A} \times \mathcal{S} \rightarrow [0,1]$ determines the probability of a transition from any  state $s \in \mathcal{S}$ to any state $s' \in \mathcal{S}$ given any possible action $a \in \mathcal{A}$. The reward function $R : \mathcal{S} \times \mathcal{A} \times \mathcal{S} \rightarrow \mathbb{R}$ defines the immediate and possibly stochastic reward that an agent would receive given that the agent executes action $a$ while in state  $s$ and it is transitioned to state $s'$, $\gamma \in [0, 1]$ represents the discount factor that balances the trade-off between immediate rewards and future rewards.

MDPs are adequate models to obtain optimal decisions in \emph{single} agent fully observable environments.\footnote{A Partially Observable Markov Decision Process (POMDP)~\cite{Cassandra:1998tz,astrom1965optimal} explicitly models environments where the agent no longer sees the true system state and instead receives an \emph{observation} (generated from the underlying system state).} Solving an MDP will yield a policy $\pi: \mathcal{S} \rightarrow \mathcal{A}$, which is a mapping from states to actions. An optimal policy $\pi^*$ is the one that maximizes the expected discounted sum of rewards. There are different techniques for solving MDPs assuming a complete description of all its elements. One of the most common techniques is the value iteration algorithm~\cite{Bellman:1957ud}, which requires a complete and accurate representation of states, actions, rewards, and transitions. However, this may be difficult to obtain in many domains. For this reason, RL algorithms often learn from experience interacting with the environment in discrete time steps.

\paragraph{Q-learning}
One of the most well known algorithms for RL is Q-learning~\cite{Watkins:1989uk}. It has been devised for stationary, single-agent, fully observable environments with discrete actions. A Q-learning agent keeps the estimate of its expected payoff starting in state $s$, taking action $a$ as $\hat Q(s,a)$. Each tabular entry $\hat Q(s,a)$ is an estimate of the corresponding optimal $Q^*$ function that maps state-action pairs to the discounted sum of future rewards starting with action $a$ at state $s$ and following the optimal policy thereafter. Each time the agent transitions from a state $s$ to a state $s'$ via action $a$ receiving payoff $r$, the $Q$ table is updated as follows:
\begin{equation}
    \hat Q(s,a) \leftarrow \hat Q(s,a) +\alpha [(r + \gamma \max_{a'} \hat Q(s',a'))- \hat Q(s,a)]
\end{equation}
with the learning rate $\alpha \in [0,1]$. Q-learning is proven \revise{to converge to} $Q^*$ if state and action spaces are discrete and finite, the sum of the learning rates goes to infinity (so that each state-action pair is visited \textit{infinitely} often) and that the sum of the squares of the learning rates is finite (which is required to show that the convergence is with probability one)~\cite{kamihigashi2015necessary,Watkins:1989uk,learning:tw,jaakkola1994convergence,szepesvari1999unified,even2003learning,szepesvari2010algorithms}. The convergence of single-step on-policy RL algorithms, i.e, SARSA ($\lambda=0$), for both decaying exploration (greedy in the limit with infinite exploration) and persistent exploration (selecting actions probabilistically according to the ranks of the Q values) was demonstrated by Singh et al.~\cite{singh2000convergence}. Furthermore, Van Seijen~\cite{VanSeijen:2009fe} has proven convergence for Expected SARSA (see Section~\ref{sec:multiagentlearning} for convergence results in multiagent domains).

\paragraph{REINFORCE (Monte Carlo policy gradient)} In contrast to value-based methods, which do not try to optimize directly over a policy space~\cite{Konda:vx}, policy gradient methods can learn parameterized policies without using intermediate value estimates. 

Policy parameters are learned by following the gradient of some performance measure with gradient descent~\cite{Sutton:1999uo}. For example, REINFORCE~\cite{williams1992simple} uses estimated return by Monte Carlo \revise{(MC)} methods with full episode trajectories to learn policy parameters $\theta$, with $\pi (a ; s,\theta) \approx \pi(a ; s)$, as follows
\begin{equation}
  \theta_{t+1} = \theta_{t} + \alpha G_{t} \frac{\nabla\pi(A_{t} ; S_{t},\theta_{t})}{\pi(A_{t} ; S_{t},\theta_{t})}
\end{equation}
 where $G_{t}$ represents the return, $\alpha$ is the learning rate, and $A_{t} \sim \pi$. A main limitation is that policy gradient methods can have high variance~\cite{Konda:vx}. 
 
 The policy gradient update can be generalized to include a comparison to an arbitrary \emph{baseline} of the state~\cite{williams1992simple}. The baseline, $b(s)$, can be any function, as long as it does not vary with the action; the baseline leaves the expected value of the update unchanged, but it can have an effect on its variance~\cite{sutton2018reinforcement}. A natural choice for the baseline is a learned state-value function, this reduces the variance, \revise{and it is bias-free if learned by MC}.\footnote{Action-dependant baselines had been proposed~\cite{liu2017action,gu2016q}, however, a recent study by Tucker et al.~\cite{tucker2018mirage} found that in many works the reason of good performance was because of bugs or errors in the code, rather than the proposed method itself.} Moreover, when using the state-value function for bootstrapping (updating the value estimate for a state from the estimated values of subsequent states) it assigns credit (reducing the variance but introducing bias), i.e., criticizes the policy's action selections. Thus, in actor-critic methods~\cite{Konda:vx}, the actor represents the policy, i.e., action-selection mechanism, whereas a critic is used for the value function learning. In the case when the critic learns a state-action function ($Q$ function) and a state value function ($V$ function), an \emph{advantage function} can be computed by subtracting state values from the state-action values~\cite{sutton2018reinforcement,Schulman:2015uk}. The advantage function indicates the relative quality of an action compared to other available actions computed from the baseline, i.e., state value function. An example of an actor-critic algorithm is Deterministic Policy Gradient (DPG)~\cite{silver2014deterministic}. In DPG~\cite{silver2014deterministic} the critic follows the standard Q-learning and the actor is updated following the gradient of the policy's performance~\cite{hafner2011reinforcement}, DPG was later extended to DRL (see Section~\ref{sec:deeprl}) and MDRL (see Section~\ref{sec:learningCooperation}). For multiagent learning settings the variance is further increased as all the agents' rewards depend on the rest of the agents, and it is formally shown that as the number of agents increase, the probability of taking a correct gradient direction decreases exponentially~\cite{Lowe:2017wb}. Recent \MDRL works addressed this high variance issue, e.g., COMA~\cite{Foerster:2018to} and MADDPG~\cite{Lowe:2017wb} (see Section~\ref{sec:learningCooperation}). %

Policy gradient methods have a clear connection with deep reinforcement learning since \emph{the policy might be represented by a neural network} whose input is a representation of the state, whose output are action selection probabilities or values for continuous control~\cite{lillicrap2015continuous}, and whose weights are the policy parameters. 

\subsection{Deep reinforcement learning}
\label{sec:deeprl}

While tabular RL methods such as Q-learning are successful in domains that do not suffer from the curse of dimensionality, there are many limitations: learning in large state spaces can be prohibitively slow, methods do not generalize (across the state space), and state representations need to be hand-specified~\cite{sutton2018reinforcement}. Function approximators tried to address those limitations, using for example, decision trees~\cite{pyeatt2001decision}, tile coding~\cite{sutton1996generalization}, radial basis functions~\cite{kretchmar1997comparison}, and locally weighted regression~\cite{boyan1995generalization} to approximate the value function. %

Similarly, these challenges can be addressed by using deep learning, i.e., neural networks~\cite{boyan1995generalization,pyeatt2001decision} as function approximators. For example, $Q(s,a ; \theta)$ can be used to approximate the state-action values with $\theta$ representing the neural network weights. This has two advantages, first, deep learning helps to generalize across states improving the sample efficiency for large state-space RL problems. Second, deep learning can be used to reduce (or eliminate) the need for manually designing features to represent state information~\cite{lecun2015deep,schmidhuber2015deep}.  

However, extending deep learning to RL problems comes with additional challenges including non-i.i.d.\ (not independently and identically distributed) data. Many supervised learning methods assume that training data \revise{is} from \revise{an} i.i.d.\ stationary distribution~\cite{bishop2006pattern,schmidhuber2015deep,riemer2018learning}. However, in RL, training data consists of highly correlated sequential agent-environment interactions, which violates the \emph{independence} condition. Moreover, RL training data distribution is non-stationary as the agent actively learns while exploring different parts of the state space, violating the condition of sampled data being \emph{identically distributed}~\cite{Mnih:2013wp}.

In practice, using function approximators in RL requires making crucial representational decisions and poor design choices can result in estimates that diverge from the optimal value function~\cite{gordon1999approximate,boyan1995generalization,baird1995residual,whiteson2006evolutionary,achiam2019divergence,vanHasselt2019triad}. In particular, function approximation, bootstrapping, and off-policy learning are considered the three main properties that when combined, can make the learning to diverge and are known as \emph{the deadly triad}~\cite{vanHasselt2019triad,sutton2018reinforcement}. Recently, some works have shown that non-linear (i.e., deep) function approximators poorly estimate the value function~\cite{fujimoto2018addressing,tucker2018mirage,ilyas2019truly} and another work found problems with Q-learning using function approximation (over/under-estimation, instability and even divergence) due to the \emph{delusional} bias: ``delusional bias occurs whenever a backed-up value estimate is derived from action
choices that are not realizable in the underlying policy class"\cite{lu2018non}.
Additionally, convergence results for reinforcement learning using function approximation are still scarce~\cite{baird1995residual,tsitsiklis1997analysis,melo2008analysis,ernst2005tree,lu2018non}; in general, stronger convergence guarantees are available for policy-gradient methods~\cite{Sutton:1999uo} than for value-based methods~\cite{sutton2018reinforcement}.

Below we mention how the existing DRL methods aim to address these challenges when briefly reviewing value-based methods, such as DQN~\cite{Mnih:2015jp}; policy gradient methods, like Proximal Policy Optimization (PPO)~\cite{Schulman:2015uk}; and actor-critic methods like Asynchronous Advantage Actor-Critic (A3C)~\cite{Jaderberg:2016uv}. We refer the reader to recent surveys on single-agent DRL~\cite{Arulkumaran:2017jn,li2017deep,franccois2018introduction} for a more detailed discussion of the literature. 

\paragraph{Value-based methods}

\begin{figure}
\centering
\includegraphics[width=\textwidth]{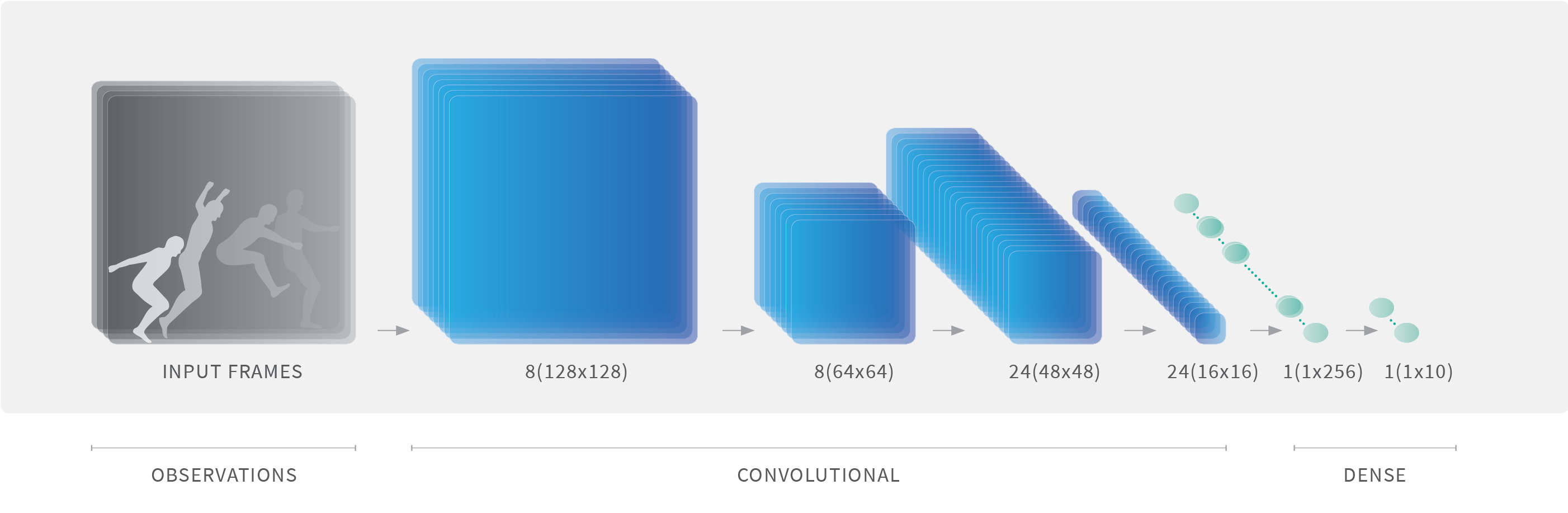}
\caption{Deep Q-Network (DQN)~\cite{Mnih:2015jp}: Inputs are four stacked frames; the network is composed of several layers: \emph{Convolutional} layers employ filters to learn features from high-dimensional data with a much smaller number of neurons and \emph{Dense} layers are fully-connected layers. The last layer represents the actions the agent can take (in this case, $10$ possible actions). Deep Recurrent Q-Network (DRQN)~\cite{Hausknecht:tq}, which extends DQN to partially observable domains~\cite{Cassandra:1998tz}, is identical to this setup except the penultimate layer ($1 \times 256$ Dense layer) is replaced with a recurrent LSTM layer~\cite{hochreiter1997long}.}
\label{fig:dqn_filter}
\end{figure}

\begin{figure}[t]
\centering
\includegraphics[width=1.0\textwidth]{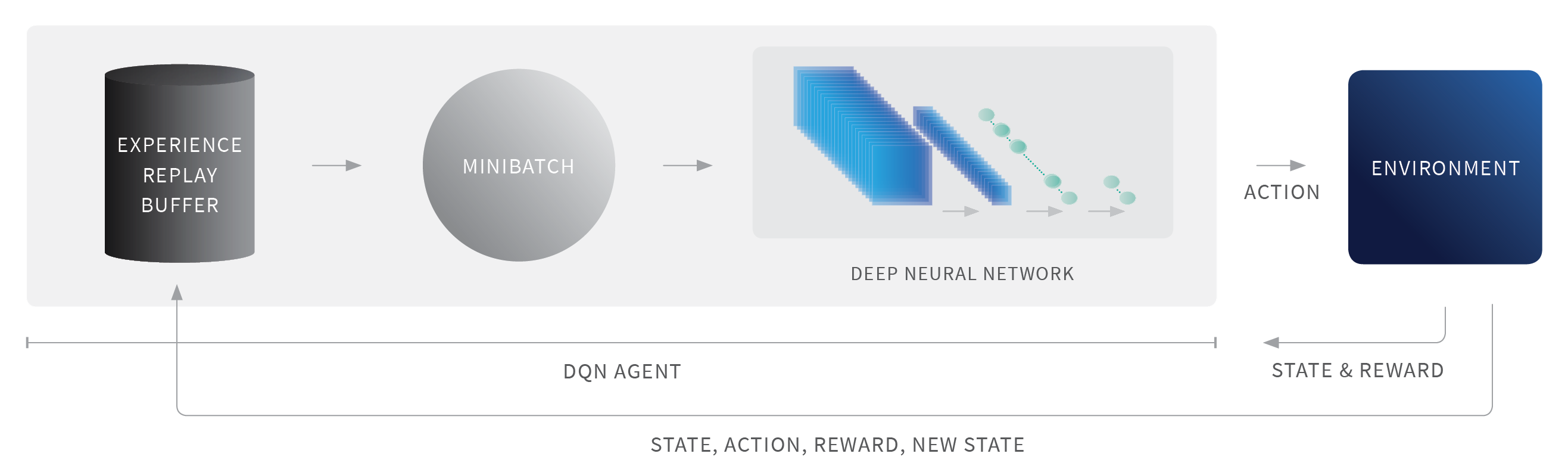}
\caption{Representation of a DQN agent that uses an experience replay buffer~\cite{lin1991programming,lin1992self} to keep $\langle s,a,r,s' \rangle$ tuples for minibatch updates. The Q-values are parameterized with a NN and a policy is obtained by selecting (greedily) over those at every timestep.}
\label{fig:dqn_diagram}
\end{figure}

The major breakthrough work combining deep learning with Q-learning was the Deep Q-Network (DQN)~\cite{Mnih:2015jp}. DQN uses a deep neural network for function approximation~\cite{riedmiller2005neural}\footnote{Before DQN, many approaches used neural networks for representing the Q-value function~\cite{crites1998elevator}, such as Neural Fitted Q-learning~\cite{riedmiller2005neural} and NEAT+Q~\cite{whiteson2006evolutionary}.} (see Figure~\ref{fig:dqn_filter}) and maintains an \emph{experience replay} (ER) buffer~\cite{lin1991programming,lin1992self} to store interactions $\langle s,a,r,s' \rangle$. DQN keeps an additional copy of neural network parameters, $\theta^-$, for the target network in addition to the $\theta$ parameters to stabilize the learning, i.e., to alleviate the non-stationary data distribution.\footnote{Double Q-learning~\cite{hasselt2010double} originally proposed keeping two $Q$ functions (estimators) to reduce the overestimation bias in RL, while still keeping the convergence guarantees, later it was extended to DRL in Double DQN~\cite{van2016deep} (see Section \ref{sec:bridging:examples}).} For each training iteration $i$, DQN minimizes the mean-squared error (MSE) between the Q-network and its target network using the loss function:

\begin{equation}
    L_{i}(\theta_i) = \E_{s, a, r, s'} [(r + \gamma max_{a'}Q(s', a';\theta_{i}^-) - Q(s, a; \theta_i))^2]
\end{equation}
where target network parameters $\theta^-$ are set to Q-network parameters $\theta$ periodically and mini-batches of $\langle s,a,r,s' \rangle$ tuples are sampled from the ER buffer, as depicted in Figure~\ref{fig:dqn_diagram}. 

The ER buffer provides stability for learning as random batches sampled from the buffer helps alleviating the problems caused by the non-i.i.d. data. However, it comes with disadvantages, such as higher memory requirements and computation per real interaction~\cite{mnih2016asynchronous}. The ER buffer is mainly used for off-policy RL methods as it can cause a mismatch between buffer content from earlier policy and from the current policy for on-policy methods~\cite{mnih2016asynchronous}.  Extending the ER buffer for the multiagent case is not trivial, see Sections~\ref{sec:learningCooperation}, \ref{sec:bridging:examples} and \ref{sec:bridging:lessons}. \revise{Recent works were designed to reduce the problem of catastrophic forgetting (this occurs when the trained neural network performs poorly on previously learned tasks due to a non-stationary training distribution~\cite{mccloskey1989catastrophic,goodfellow2013empirical}) and the ER buffer, in DRL~\cite{isele2018selective} and MDRL~\cite{palmer2018negative}.}

DQN has been extended in many ways, for example, by using double estimators~\cite{hasselt2010double} to reduce the overestimation bias with Double DQN~\cite{van2016deep} (see Section \ref{sec:bridging:examples}) and
by decomposing the Q-function with a \emph{dueling}-DQN architecture~\cite{wang2015dueling}, where two streams are learned, one estimates state values and another one advantages, those are combined \revise{in} the final layer to form $Q$ values (this method improved over Double DQN). 

 In practice, DQN is trained using an input of four stacked frames (last four frames the agent has encountered). If a game requires a memory of more than four frames it will appear non-Markovian to DQN because \revise{the future game states (and rewards) do not depend only on the input (four frames) but rather on the history~\cite{Hauskrecht:2000uh}.} Thus, DQN's performance declines when given incomplete state observations (e.g., one input frame) since DQN assumes full state observability.

Real-world tasks often feature incomplete and noisy state information resulting from \emph{partial observability} (see Section~\ref{sec:rl}). Deep Recurrent Q-Networks (DRQN)~\cite{Hausknecht:tq} proposed using \emph{recurrent neural networks}, in particular, Long Short-Term Memory (LSTMs) cells~\cite{hochreiter1997long} in DQN, for this setting. Consider the architecture in Figure~\ref{fig:dqn_filter} with the first dense layer after convolution replaced by a layer of LSTM cells. \revise{With} this addition, DRQN has memory capacity so that it can even work with only one input frame rather than a stacked input of consecutive frames. This idea has been extended to \MDRL, see Figure~\ref{fig:DPIRQN} and Section~\ref{sec:bridging:lessons}. \revise{There are also other approaches to deal with partial observability such as finite state controllers~\cite{meuleau1999learning} (where action selection is performed according to the complete observation history) and using an initiation set of options  conditioned on the previously employed option~\cite{steckelmacher2018reinforcement}.}

\paragraph{Policy gradient methods}

For many tasks, particularly for physical control, the action space is continuous and high dimensional where DQN is not suitable. Deep Deterministic Policy Gradient (DDPG)~\cite{lillicrap2015continuous} is a model-free off-policy actor-critic algorithm for such domains, based on the DPG algorithm~\cite{silver2014deterministic} (see Section~\ref{sec:rl}). Additionally, it proposes a new method for updating the networks, i.e., the target network parameters slowly change (this could also be applicable to DQN), in contrast to the hard reset (direct weight copy) used in DQN. Given the off-policy nature, DDPG generates exploratory behavior by adding sampled noise from some noise processes to its actor policy. The authors also used batch normalization~\cite{ioffe2015batch} to ensure generalization across many different tasks without performing manual normalizations. However, note that other works have shown batch normalization can cause divergence in DRL~\cite{van2016coordinated,salimans2016weight}.

\begin{figure}[t]
\centering
\includegraphics[width=\textwidth]{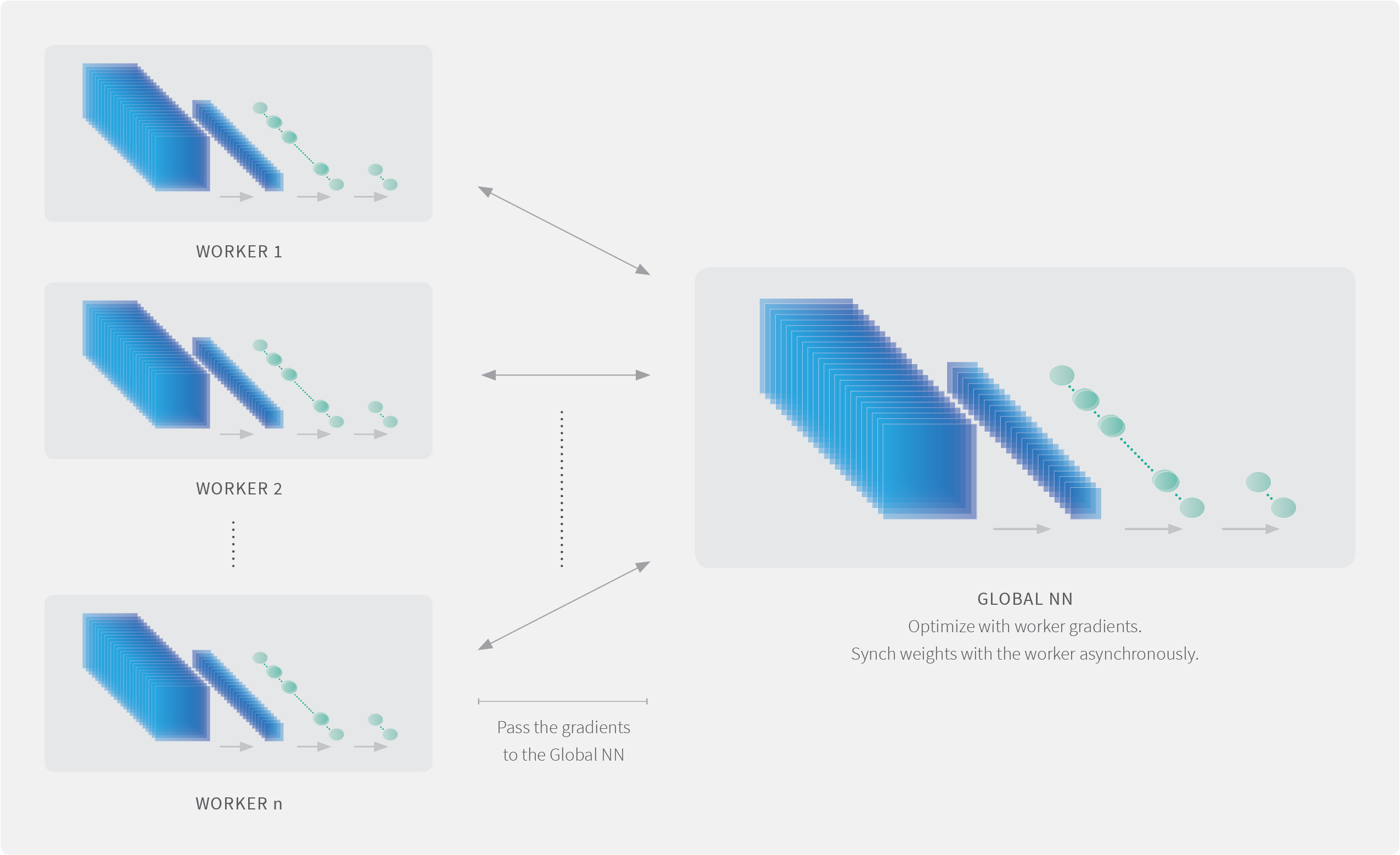}
\caption{Asynchronous Advantage Actor-Critic (A3C) employs multiple (CPUs) workers without needing an ER buffer. Each worker has its own NN and independently interacts with the environment to compute the loss and gradients. Workers then pass computed gradients to the global NN that optimizes the parameters and synchronizes with the worker \emph{asynchronously}. This distributed system is designed for single-agent deep RL. \revise{Compared to different DQN variants, A3C obtains better performance on a variety of Atari games using substantially less training time with multiple CPU cores of standard laptops without a GPU~\cite{mnih2016asynchronous}. However, we note that more recent approaches use both multiple CPU cores for more efficient training data generation and GPUs for more efficient learning.}}
\label{fig:a3c}
\end{figure}

Asynchronous Advantage Actor-Critic (A3C)~\cite{mnih2016asynchronous}
is an algorithm that employs a \emph{parallelized} asynchronous training scheme (using multiple CPU threads) for efficiency. It is an on-policy RL method that does not use an experience replay buffer. A3C allows multiple workers to simultaneously interact with the environment and compute gradients locally. All the workers pass their computed local gradients to a global NN which performs the optimization and synchronizes with the workers asynchronously (see Figure~\ref{fig:a3c}). There is also the Advantage Actor-Critic (A2C) method~\cite{a2c2017} that combines all the gradients from all the workers to update the global NN \emph{synchronously}. The loss function for A3C is composed of two terms: policy loss (actor), $\mathcal{L}_{\pi}$, and value loss (critic), $\mathcal{L}_{v}$. A3C parameters are updated using the \emph{advantage} function $A(s_t, a_t; \theta_v)=Q(s,a)-V(s)$, commonly used to reduce variance (see Section~\ref{sec:rl}). An entropy loss for the policy, $H(\pi)$, is also commonly added, which helps to improve exploration by discouraging premature convergence to suboptimal deterministic policies~\cite{mnih2016asynchronous}.
Thus, the loss function is given by: 
    $\mathcal{L}_{\text{A3C}} = \lambda_v  \mathcal{L}_{v} + \lambda_{\pi} \mathcal{L}_{\pi} - \lambda_{H} \mathbb{E}_{s \sim \pi} [H(\pi(s, \cdot, \theta)]$ 
with $\lambda_{v}, \lambda_{\pi},$ \revise{and} $\lambda_{H}$, being weighting terms on the individual loss components.  
\revise{Wang et al.}~\cite{wang2016sample} took A3C's framework but used off-policy learning \revise{to create the Actor-critic with experience replay (ACER) algorithm.} \revise{Gu et al.~\cite{gu2017interpolated} introduced the} Interpolated Policy Gradient (IPG) \revise{algorithm and} showed a connection between ACER and DDPG: they are a pair of reparametrization terms (they are special cases of IPG) when they are put under the same stochastic policy setting, and when the policy is deterministic they collapse into DDPG.

\revise{Jaderberg et al.~\cite{Jaderberg:2016uv} built the} Unsupervised Reinforcement and Auxiliary Learning (UNREAL) framework \revise{on top of} A3C and introduced unsupervised \emph{auxiliary tasks} (e.g., reward prediction) to speed up the learning process. Auxiliary tasks in general are not used for anything other than shaping the features of the agent, i.e., facilitating and regularizing the representation learning process~\cite{shelhamer2016loss,bellemare2019geometric}; their formalization in RL is related to the concept of \emph{general value functions}~\cite{sutton2018reinforcement,sutton2011horde}.
The UNREAL framework optimizes a combined loss function $\mathcal{L}_{\text{UNREAL}} \approx \mathcal{L}_{\text{A3C}} + \sum_i \lambda_{AT_i} \mathcal{L}_{AT_i} $, that combines the A3C loss, $\mathcal{L}_{\text{A3C}}$, together with auxiliary task losses $\mathcal{L}_{AT_i}$, where $\lambda_{AT_i}$ are weight terms (see Section~\ref{sec:bridging:examples} for use of auxiliary tasks in \MDRL). 
In contrast to A3C, UNREAL uses a prioritized ER buffer, in which transitions with positive reward are given higher probability of being sampled.
This approach can be viewed as a simple form of prioritized replay~\cite{Schaul:2015ty}, which was in turn inspired by model-based RL algorithms like prioritized sweeping~\cite{moore1993prioritized,andre1998generalized}. 

Another distributed architecture is the Importance Weighted Actor-Learner Architecture (IMPALA)~\cite{espeholt2018impala}. Unlike
A3C or UNREAL, IMPALA actors communicate \emph{trajectories
of experience} (sequences of states, actions, and rewards) to a centralized learner, thus IMPALA decouples acting from learning.

Trust Region Policy Optimization (TRPO)~\cite{Schulman:2015uk} and Proximal Policy Optimization (PPO)~\cite{Schulman:2017vq} are recently proposed policy gradient algorithms where the latter represents the state-of-the art with advantages such as being simpler to implement and having better empirical sample complexity. Interestingly, a recent work~\cite{ilyas2019truly} studying PPO and TRPO arrived at the surprising conclusion that these methods often deviate from what the theoretical framework would predict: gradient estimates are poorly correlated with the true gradient and value networks tend to produce inaccurate predictions for the true value function. Compared to vanilla policy gradient algorithms, PPO prevents abrupt changes in policies during training through the loss function, similar to early work by Kakade~\cite{kakade2002natural}.  Another advantage of PPO is that it can be used in a distributed fashion, i.e, Distributed PPO (DPPO)~\cite{Heess:2017ts}. Note that \emph{distributed approaches} like DPPO or A3C use \revise{ parallelization only to improve the learning by more efficient training data generation through multiple CPU cores} for single agent DRL and they should not be considered multiagent approaches (except \revise{for} recent work which tries to exploit this parallelization in a multiagent environment~\cite{bacchiani2019microscopic}).

Lastly, there's a connection between policy gradient algorithms and Q-learning~\cite{schulman2017equivalence} within the framework of entropy-regularized reinforcement learning~\cite{haarnoja2017reinforcement} where the value and $Q$ functions are slightly altered to consider the entropy of the policy. In this vein, Soft Actor-Critic (SAC)~\cite{haarnoja2018soft} is a recent algorithm that concurrently learns a stochastic policy, two Q-functions (taking inspiration from Double Q-learning) and a value function. SAC alternates between collecting experience with the current policy and updating from batches sampled from the ER buffer.

We have reviewed recent algorithms in DRL, while the list is not exhaustive, it provides an overview of the different state-of-art techniques and algorithms which will become useful while describing the \MDRL techniques in the next section. 
 
\section{Multiagent Deep Reinforcement Learning (\MDRL)}
\label{sec:deepmal}

First, we briefly introduce the general framework on multiagent learning and then we dive into the categories and the research on \MDRL.

\subsection{Multiagent Learning}
 \label{sec:multiagentlearning}

Learning in a multiagent environment is inherently more complex than in the single-agent case, as agents interact at the same time with environment and potentially with each other~\cite{Busoniu:2008bo}. The \emph{independent} learners, a.k.a. \emph{decentralized} learners approach~\cite{Tan:1993em} directly uses single-agent algorithms in the multi-agent setting despite the underlying assumptions of these algorithms being violated (each agent independently learns its own policy, treating other agents as part of the environment). In particular the \emph{Markov property} (the future dynamics, transitions, and rewards depend only on the current state) becomes invalid since the environment is no longer stationary~\cite{Tuyls:2012up,nowe2012game,laurent2011world}. This approach ignores the multiagent nature of the setting entirely and it can fail when an opponent adapts or learns, for example, based on the past history of interactions~\cite{Shoham:2007vw}. Despite the lack of guarantees, independent learners have been used in practice, providing advantages with regards to scalability while often achieving good results~\cite{Matignon:2012bj}.

To understand why multiagent domains are non-stationary from agents' local perspectives, consider a simple stochastic (also known as Markov) game $(\mathcal{S},\mathcal{N},\mathcal{A},\mathcal{T},\mathcal{R})$, which can be seen as an extension of an MDP to multiple agents~\cite{Littman:1994ta,littman2001value}. One key distinction is that the transition, $\mathcal{T}$, and reward function, $\mathcal{R}$, depend on the actions $\mathcal{A}= A_1\times ... \times A_\mathcal{N}$ of all, $\mathcal{N}$, agents, this means,  $\mathcal{R}=R_1\times ... \times R_\mathcal{N}$ and  $\mathcal{T}=\mathcal{S} \times A_1\times ... \times A_\mathcal{N}$. 

Given a learning agent $i$ and using the common shorthand notation $\bm{-i} = \mathcal{N} \setminus \{ i \}$ for the set of opponents, the value function now depends on the joint action $\bm{a} = (a_i, \bm{a_{-i}})$, and the joint policy $\bm{\pi}(s, \bm{a}) = \prod_j \pi_j (s, a_j) $:\footnote{In this setting each agent independently executes a policy, however, there are other cases where this does not hold, for example when agents have a coordinated exploration strategy.}

\begin{equation}
\label{eqn:bellmanMAS}
V^{\bm{\pi}}_{i}(s)=  \sum_{\bm{a} \in \mathcal{A}} \bm{\pi}(s,\bm{a})  \sum_{s' \in \mathcal{S}}  \mathcal{T}(s,a_i,\bm{a_{-i}},s') [R_i(s,a_i,\bm{a_{-i}},s') + \gamma V_{i}(s')].
\end{equation}
Consequently, the optimal policy is %
dependent on the other agents' policies,
\begin{equation}
    \begin{aligned}
        &\pi_i^*(s,a_i,\bm{\pi_{-i}}) = %
        \argmax_{\pi_i} V^{(\pi_i, \bm{\pi_{-i}})}_{i}(s) =\\
        & \argmax_{\pi_i} \sum_{\bm{a} \in \mathcal{A}} \pi_i(s, a_i) \bm{\pi_{-i}}(s,\bm{a_{-i}})  \sum_{s' \in \mathcal{S}}  \mathcal{T}(s,a_i,\bm{a_{-i}},s') [R_i(s,a_i,\bm{a_{-i}},s') + \gamma V^{(\pi_i, \bm{\pi_{-i}})}_{i}(s')].
    \end{aligned}
\end{equation}
Specifically, the opponents' joint policy $\bm{\pi_{-i}}(s,\bm{a_{-i}})$ can be non-stationary, i.e., changes as
the opponents' policies change over time, for example with learning opponents.%

\paragraph{Convergence results} 

Littman~\cite{littman2001value} studied convergence properties of reinforcement learning \revise{joint action agents~\cite{Claus:1998tb}} in Markov games  with the following conclusions: in adversarial environments (zero-sum games) an optimal play can be guaranteed against an arbitrary opponent, i.e., Minimax Q-learning~\cite{Littman:1994ta}. \revise{In} coordination environments (e.g., in cooperative games all agents share the same reward function), strong assumptions need be made about other agents to guarantee convergence to optimal behavior~\cite{littman2001value}, e.g., Nash Q-learning~\cite{Hu:2003wb} and Friend-or-Foe Q-learning~\cite{Littman:2001vc}. In other types of environments no value-based RL algorithms with guaranteed convergence properties are known~\cite{littman2001value}. 

Recent work on \MDRL have addressed scalability and have focused significantly less on convergence
guarantees, with few exceptions~\cite{song2019convergence,balduzzi2018mechanics,perolat2018actor,bono2018cooperative}. One notable work has shown a connection between update rules for actor-critic algorithms for multiagent partially observable settings and (counterfactual) regret minimization:\footnote{
Counterfactual regret minimization is a technique for solving large games based on regret minimization~\cite{neller2013introduction,zinkevich2008regret}
due to a well-known connection between regret and Nash equilibria~\cite{blum2007learning}. It has been one of the reasons of successes in Poker~\cite{Moravcik:2017jb,brown2018superhuman}.} the advantage values are scaled counterfactual regrets. This lead to new convergence properties of independent RL algorithms in zero-sum games with imperfect information~\cite{srinivasan2018actor}.  The result is also used to support policy gradient optimization against worst-case opponents, in a new algorithm called Exploitability Descent~\cite{lockhart2019computing}.\footnote{This algorithm is similar to CFR-BR~\cite{johanson2012finding} and has the main advantage that the current policy convergences rather than the average policy, so there is no need to learn the average strategy, which requires large reservoir buffers or many past networks.}

We refer the interested reader to seminal works about convergence in multiagent domains~\cite{kalai1993rational,sandholm1996multiagent,singh2000nash,banerjee2003adaptive,Greenwald:2003uy,bowling2000convergence,Bowling:2005vi,zinkevich2006cyclic,Bowling:2002vva,Kaisers:2011uf,wunder2010classes}. Note that instead of convergence, some MAL algorithms have proved learning a best response against classes of opponents~\cite{Tesauro:2003wq,Weinberg:2004wj,Chakraborty:2013ii}.

There are other common problems in MAL, including action shadowing~\cite{Wei:2016wq,Fulda:2006tu}, the curse of dimensionality~\cite{Busoniu:2008bo}, and multiagent credit assignment~\cite{Agogino:2004ug}. Describing each problem is out of the scope of this survey. However, we refer the interested reader to excellent resources on general MAL~\cite{Tuyls:2012up,Weiss:2013ue,multiagenttutorial}, as well as surveys in specific areas: game theory and multiagent reinforcement learning~\cite{Busoniu:2008bo,nowe2012game}, cooperative scenarios~\cite{Panait:2005wj,Matignon:2012bj}, evolutionary dynamics of multiagent learning~\cite{Bloembergen:2015ei}, learning in non-stationary environments~\cite{HernandezLeal:2017wd}, agents modeling agents~\cite{Albrecht:2018dp}, and transfer learning in multiagent RL~\cite{da2019survey}.

\subsection{\MDRL categorization}
\label{sec:categorization}

In \revise{Section~\ref{sec:deeprl}} we outlined some recent works in single-agent DRL since an exhaustive list is out of the scope of this article. This explosion of works has led DRL to be extended and combined with other techniques~\cite{Arulkumaran:2017jn,li2017deep,papoudakis2019dealing}. One natural extension to DRL is to test whether these approaches could be applied in a multiagent environment.

We analyzed the most recent works (that are not covered by previous MAL surveys~\cite{HernandezLeal:2017wd,Albrecht:2018dp} and we do not consider genetic algorithms or swarm intelligence in this survey) that have a clear connection with \MDRL.  We propose 4 categories which take inspiration from previous surveys~\cite{Stone:2000ww,Busoniu:2008bo,Panait:2005wj,Albrecht:2018dp} and that conveniently describe and represent current works. Note that some of these works fit into more than one category (they are not mutually exclusive), therefore their summaries are presented in all applicable   Tables~\ref{tab:emergent}-\ref{tab:modeling}, however, for the ease of exposition when describing them in the text we only do so in one category. Additionally, for each work we present its learning type, either a value-based method (e.g., DQN) or a policy gradient method (e.g., actor-critic); also, we mention if the setting is evaluated in a fully cooperative, fully competitive or mixed environment (both cooperative and competitive).

\begin{itemize}
\item \emph{Analysis of emergent behaviors}. These works, in general, do not propose learning algorithms --- their main focus is to analyze and evaluate DRL algorithms, e.g., DQN~\cite{Tampuu:2017fc,Leibo:2017wi,Raghu:2018tg}, PPO~\cite{Bansal:2017wl,Raghu:2018tg} and others~\cite{leibo2018malthusian,Raghu:2018tg,mordatch2017emergence}, in a multiagent environment. In this category we found works which analyze behaviors in the three major settings: cooperative, competitive and mixed scenarios; see Section~\ref{sec:emergentBehaviors} and Table~\ref{tab:emergent}.

\item \emph{Learning communication}~\cite{Lazaridou:2016un,mordatch2017emergence,Foerster:2016ud,Sukhbaatar:2016tk,Peng:vd,pesce2019improving}. These works explore a sub-area in which agents can share information with communication protocols, for example through direct messages~\cite{Foerster:2016ud} or via a shared memory~\cite{pesce2019improving}. This area is attracting attention \revise{and it had not been explored} much in the MAL literature. See Section~\ref{sec:learningCommunication} and Table~\ref{tab:communication}.

\item \emph{Learning cooperation}. While learning to communicate is an emerging area, fostering cooperation in learning agents has a long history of research in MAL~\cite{Panait:2005wj,Matignon:2012bj}. In this category the analyzed works are evaluated in either cooperative or mixed settings. Some works in this category take inspiration from MAL (e.g., leniency, hysteresis, and difference rewards concepts) and extend them to the \MDRL setting~\cite{Palmer:2018wv,Omidshafiei:2017uk,Foerster:2017uq}. A notable exception~\cite{Foerster:2017ti} takes a key component from RL (i.e., experience replay buffer) and adapts it for MDRL. See Section~\ref{sec:learningCooperation} and Table~\ref{tab:cooperation}.

\item \emph{Agents modeling agents}. Albrecht and Stone~\cite{Albrecht:2018dp} presented a thorough survey in this topic and we have found many works that fit into this category in the \MDRL setting, some taking inspiration from DRL~\cite{He:2016up,Raileanu:2018tb,Hong:2018vp}, and others from MAL~\cite{Lanctot:vn,Heinrich:69lSVXjW,Foerster:2018to,Rabinowitz:2018uf,Yang:4pJjEXeO}. Modeling agents is helpful not only to cooperate, but also for modeling opponents~\cite{Lanctot:vn,He:2016up,Hong:2018vp,Heinrich:69lSVXjW}, inferring goals~\cite{Raileanu:2018tb}, and accounting for the learning behavior of other agents~\cite{Foerster:2018to}. In this category the analyzed algorithms present their results in either a competitive setting or a mixed one (cooperative and competitive). See Section~\ref{sec:agentsModeling} and Table~\ref{tab:modeling}.

\end{itemize}

In the rest of this section we describe each category along with the summaries of related works.

\captionsetup{belowskip=8pt,aboveskip=0pt}
\begin{table}
\footnotesize
\centering
\caption{These papers analyze \emph{emergent behaviors} in \MDRL. Learning type \revise{is} either value-based (VB) or policy gradient (PG). Setting where experiments were performed: cooperative (CO), competitive (CMP) or mixed. A detailed description is given in Section~\ref{sec:emergentBehaviors}.}
\label{tab:emergent}
\begin{tabular}{@{}p{3.1cm}p{8.0cm}cc@{}} \toprule
Work &  Summary  & Learning & Setting \\ \hline
Tampuu et al.~\cite{Tampuu:2017fc} &  Train DQN agents to play Pong.& VB & CO\&CMP \\
Leibo et al.~\cite{Leibo:2017wi} &  Train DQN agents to play sequential social dilemmas. & VB & Mixed\\
Lerer and Peysakhovich~\cite{lerer2017maintaining} & Propose DRL agents able to cooperate in social dilemmas. & VB&  Mixed \\
Leibo et al.~\cite{leibo2018malthusian} & Propose Malthusian reinforcement learning which extends self-play to population dynamics. & VB & Mixed\\
Bansal et al.~\cite{Bansal:2017wl} &  Train PPO agents in competitive MuJoCo scenarios. & PG  & CMP\\
Raghu et al.~\cite{Raghu:2018tg} &  Train PPO, A3C, and DQN agents in attacker-defender games. & VB, PG & CMP\\
Lazaridou et al.~\cite{Lazaridou:2016un} & Train agents represented with NN to learn a communication language. & PG & CO \\
Mordatch and Abbeel~\cite{mordatch2017emergence} & Learn communication with an end-to-end differentiable model to train with backpropagation. &  PG & CO \\
\bottomrule
\end{tabular}

\caption{These papers propose algorithms for \emph{learning communication}. Learning type \revise{is} either value-based (VB) or policy gradient (PG). Setting were experiments were performed: cooperative (CO) or mixed. A more detailed description is given in Section~\ref{sec:learningCommunication}.}
\label{tab:communication}
\begin{tabular}{@{}p{3.1cm}p{8.2cm}cc@{}} \toprule
Algorithm & Summary & Learning & Setting\\ \hline
Lazaridou et al.~\cite{Lazaridou:2016un}   & Train agents represented with NN to learn a communication language. & PG & CO\\
Mordatch and Abbeel~\cite{mordatch2017emergence} & Learn communication with an end-to-end differentiable model to train with backpropagation. & PG & CO\\
RIAL~\cite{Foerster:2016ud}   & Use a single network (parameter sharing) to train agents that take environmental and communication actions. & VB & CO\\
DIAL~\cite{Foerster:2016ud}  & Use gradient sharing during learning and communication actions during execution. & VB  & CO\\
CommNet~\cite{Sukhbaatar:2016tk}   &  Use a continuous vector channel for communication on a single network.& PG & CO \\
BiCNet~\cite{Peng:vd} & Use the actor-critic paradigm where communication occurs in the latent space. & PG & Mixed \\
MD-MADDPG~\cite{pesce2019improving}  & Use of a shared memory as a means to multiagent communication. & PG & CO \\
MADDPG-MD~\cite{kim2019message} & Extend dropout technique to robustify communication when applied in multiagent scenarios with direct communication. & PG  & CO\\
\bottomrule
\end{tabular}
\end{table}

\captionsetup{belowskip=8pt,aboveskip=0pt}
\begin{table}
\footnotesize
\centering
\caption{These papers aim to \emph{learn cooperation}. Learning type \revise{is} either value-based (VB) or policy gradient (PG). Setting where experiments were performed: cooperative (CO), competitive (CMP) or mixed. A more detailed description is given in Section~\ref{sec:learningCooperation}.}
\label{tab:cooperation}
\centering
\begin{tabular}{@{}p{3.2cm}p{8.0cm}cc@{}} \toprule
Algorithm   & Summary & Learning & Setting\\ \hline
Lerer and Peysakhovich~\cite{lerer2017maintaining} & Propose DRL agents able to cooperate in social dilemmas. & VB & Mixed\\
MD-MADDPG~\cite{pesce2019improving}  &Use of a shared memory as a means to multiagent communication.& PG & CO\\
MADDPG-MD~\cite{kim2019message} & Extend dropout technique to robustify communication when applied in multiagent scenarios with direct communication. & PG& CO\\
RIAL~\cite{Foerster:2016ud} & Use a single network (parameter sharing) to train agents that take environmental and communication actions. & VB& CO\\
DIAL~\cite{Foerster:2016ud}  & Use gradient sharing during learning and communication actions during execution. & VB& CO\\
DCH/PSRO~\cite{Lanctot:vn}  &Policies can overfit to opponents: better compute approximate best responses to a mixture of policies. & VB & CO \& CMP \\
Fingerprints~\cite{Foerster:2017ti}  & Deal with ER problems in \MDRL by conditioning the value function on a fingerprint that disambiguates the age of the sampled data. & VB & CO\\
Lenient-DQN~\cite{Palmer:2018wv}  & Achieve cooperation by leniency, optimism in the value function by forgiving suboptimal (low-rewards) actions. & VB & CO\\
Hysteretic-DRQN~\cite{Omidshafiei:2017uk}  & Achieve cooperation by using two learning rates, depending on the updated values together with multitask learning via policy distillation. & VB & CO\\
WDDQN~\cite{zheng2018weighted}  & Achieve cooperation by leniency, weighted double estimators, and a modified prioritized experience replay buffer. & VB & CO\\
FTW~\cite{Jaderberg:2018wx}  & Agents act in a mixed environment (composed of teammates and opponents), it proposes a two-level architecture and population-based learning. & PG & Mixed \\
VDN~\cite{Sunehag:2018vua} & Decompose the team action-value function into pieces across agents, where the pieces can be easily added. & VB & Mixed\\
QMIX~\cite{Rashid:2018wy} & Decompose the team action-value function together with a mixing network that can recombine them. & VB & Mixed \\
COMA~\cite{Foerster:2017uq} &  Use a centralized critic and a counter-factual advantage function based on solving the multiagent credit assignment. & PG& Mixed\\
PS-DQN, PS-TRPO, PS-A3C~\cite{10.1007/978-3-319-71682-4_5}  & Propose parameter sharing for learning cooperative tasks. & VB, PG & CO\\
MADDPG~\cite{Lowe:2017wb} & Use an actor-critic approach where the critic is augmented with information from other agents, the actions of all agents. & PG & Mixed
\\\bottomrule
\end{tabular}
\end{table}

\captionsetup{belowskip=8pt,aboveskip=0pt}
\begin{table}
\footnotesize
\centering
\caption{These papers consider \emph{agents modeling agents}. Learning type \revise{is} either value-based (VB) or policy gradient (PG). Setting where experiments were performed: cooperative (CO), competitive (CMP) or mixed. A more detailed description is given in Section~\ref{sec:agentsModeling}.}
\label{tab:modeling}
\centering
\begin{tabular}{@{}p{2.7cm}p{8.3cm}cc@{}} \toprule
Algorithm &  Summary & Learning & Setting\\ \hline
MADDPG~\cite{Lowe:2017wb}& Use an actor-critic approach where the critic is augmented with information from other agents, the actions of all agents. & PG & Mixed\\
DRON~\cite{He:2016up} & Have a network to infer the opponent behavior together with the standard DQN architecture. & VB & Mixed\\
DPIQN, DPIRQN~\cite{Hong:2018vp}  &  Learn policy features from raw observations that represent high-level opponent behaviors via auxiliary tasks. & VB & Mixed\\
SOM~\cite{Raileanu:2018tb} & Assume the reward function depends on a hidden goal of both agents and then use an agent's own policy to infer the goal of the other agent. & PG  & Mixed\\
NFSP~\cite{Heinrich:69lSVXjW}&  Compute approximate Nash equilibria via self-play and two neural networks. & VB & CMP\\
PSRO/DCH~\cite{Lanctot:vn}&  Policies can overfit to opponents: better compute approximate best responses to a mixture of policies. & PG  & CO \& CMP \\
M3DDPG~\cite{li2019robust}  & Extend MADDPG with minimax objective to robustify the learned policy. & PG & Mixed\\
LOLA~\cite{Foerster:2018to}&  Use a learning rule where the agent accounts for the parameter update of other agents to maximize its own reward. & PG & Mixed\\
ToMnet~\cite{Rabinowitz:2018uf}&  Use an architecture for end-to-end learning and inference of diverse opponent types. & PG & Mixed\\
Deep Bayes-ToMoP~\cite{Yang:4pJjEXeO}&  Best respond to opponents using Bayesian policy reuse, theory of mind, and deep networks. & VB & CMP \\
Deep BPR+\cite{zheng2018deep} & Bayesian policy reuse and policy distillation to quickly best respond to opponents. & VB & CO \& CMP\\
\bottomrule
\end{tabular}
\end{table}

\subsection{Emergent behaviors}
\label{sec:emergentBehaviors}

Some recent works have analyzed the previously mentioned \emph{independent} DRL agents (see Section~\ref{sec:multiagentlearning}) from the perspective of types of emerging behaviors (e.g., cooperative or competitive).

One of the earliest \MDRL works is by Tampuu et al.~\cite{Tampuu:2017fc}, which had two independent DQN learning agents to play the Atari Pong game. Their focus was to adapt the reward function for the learning agents, which resulted in either cooperative or competitive emergent behaviors.

Leibo et al.~\cite{Leibo:2017wi} meanwhile studied independent DQNs in the context of \emph{sequential social dilemmas}: a Markov game that satisfies certain inequalities~\cite{Leibo:2017wi}. The focus of this work was to highlight that cooperative or competitive behaviors exist not only as discrete (atomic) actions, but they are temporally extended (over policies). In the related setting of one shot Markov social dilemmas, Lerer and Peysakhovich~\cite{lerer2017maintaining} extended the famous Tit-for-Tat (TFT)\footnote{TFT originated in an iterated prisoner's dilemma tournament and later inspired different strategies in MAL~\cite{Powers:2005ws}, its generalization, Godfather, is a representative of \emph{leader strategies}~\cite{Littman:2001vaa}.} strategy~\cite{Axelrod:1981vw} for DRL (using function approximators) and showed (theoretically and experimentally) that such agents can maintain cooperation. To construct the agents they used self-play and two reward schemes: selfish and cooperative. Previously, different MAL algorithms \revise{were designed} to foster cooperation in social dilemmas with Q-learning agents~\cite{MunozdeCote:2006du,stimpson2003learning}.

 {Self-play is a useful concept for learning algorithms (e.g., fictitious play~\cite{Brown:1951vc}) since under certain classes of games it can guarantee convergence\footnote{The average strategy profile of fictitious players converges to a Nash equilibrium in certain classes of games,
e.g., two-player zero-sum and potential games~\cite{monderer1996fictitious}.} and it has been used as a standard technique in previous RL and MAL works~\cite{tesauro1995temporal,Silver:2016hl,Bowling:2015ub}. Despite its common usage self-play can be brittle to forgetting past knowledge~\cite{leibo2019autocurricula,Lanctot:vn,samothrakis2013coevolving} (see Section~\ref{sec:bridging:questions} for a note on the role of self-play as \revise{an} open question in \MDRL). To overcome this issue, Leibo et al.~\cite{leibo2018malthusian} proposed Malthusian reinforcement learning as an extension of self-play to population dynamics. The approach can be thought of as community coevolution and has been shown to produce better results (avoiding local optima) than independent agents with intrinsic motivation~\cite{bellemare2016unifying}. A limitation  of this work is that it does not place itself within the state of the art in evolutionary and genetic algorithms. Evolutionary strategies have been employed for solving reinforcement learning problems~\cite{moriarty1999evolutionary} and for evolving function approximators~\cite{whiteson2006evolutionary}. Similarly, they have been used multiagent scenarios to compute approximate Nash equilibria~\cite{oliehoek2006parallel} and as metaheuristic optimization algorithms~\cite{bull1995evolution,bull1998evolutionary,Panait:2005wj,iba1996emergent}.

Bansal et al.~\cite{Bansal:2017wl} explored the emergent behaviors in competitive scenarios using the MuJoCo simulator~\cite{Todorov:2012vi}. They trained independent learning agents with PPO \revise{and incorporated} two main modifications to deal with the MAL nature of the problem. First, they used \emph{exploration rewards}~\cite{gullapalli1992shaping} which are dense rewards that allow agents to learn basic (non-competitive) behaviors --- this type of reward is annealed through time giving more weight to the environmental (competitive) reward. Exploration rewards come from early work in robotics~\cite{mahadevan1992automatic} and single-agent RL~\cite{konidaris2006autonomous}, and their goal is to provide dense feedback for the learning algorithm to improve sample efficiency (Ng et al.~\cite{Ng:1999vl} studied the theoretical conditions under which modifications of the reward function of an MDP preserve the optimal policy). For multiagent scenarios, these dense rewards help agents in the beginning phase of the training to learn basic non-competitive skills, increasing the probability of random actions from the agent yielding a positive reward. The second contribution was \emph{opponent sampling} which maintains a pool of older versions of the opponent to sample from, in contrast to using the most recent version.

Raghu et al.~\cite{Raghu:2018tg} investigated how DRL algorithms (DQN, A2C, and PPO) performed in a family of two-player zero-sum games with tunable complexity, called Erdos-Selfridge-Spencer games~\cite{erdos1973combinatorial,spencer1994randomization}. Their reasoning is threefold: (i) these games provide a parameterized family of environments where (ii) optimal behavior can be completely characterized, and (iii) support multiagent play. Their work showed that algorithms can exhibit wide variation in performance as the algorithms are tuned to the game's difficulty.

Lazaridou et al.~\cite{Lazaridou:2016un} proposed a framework for language learning that relies on multiagent communication. The agents, represented by (feed-forward) neural networks, need to develop an \emph{emergent language} to solve a task. The task is formalized as a \emph{signaling game}~\cite{Fudenberg:1991vw} in which two agents, a sender and a receiver, obtain a pair of images. The sender is told one of them is the target and is allowed to send a message (from a fixed vocabulary) to the receiver. Only when the receiver identifies the target image do both agents receive a positive reward. The results show that agents can coordinate for the experimented visual-based domain. To analyze the semantic properties\footnote{The vocabulary that agents use was arbitrary and had no initial meaning. To understand its emerging semantics they looked at the relationship between symbols and the sets of images they referred to~\cite{Lazaridou:2016un}.} of the learned communication protocol they looked whether symbol usage reflects the semantics of the visual space, and that despite some variation, many high level objects groups correspond to the same learned symbols using a t-SNE~\cite{maaten2008visualizing} based analysis (t-SNE is a visualization technique for high-dimensional data and it has also been used to better understand the behavior of trained DRL agents~\cite{zahavy2016graying,beeching2019deep}).
 A key objective of this work was to determine if the agent's language could be human-interpretable. To achieve this, learned symbols were grounded with natural language by extending the signaling game with a supervised image labelling task (the sender will be encouraged to use conventional names, making communication more transparent to humans). To measure the interpretability of the extended game, a crowdsourced survey was performed, and in essence, the trained agent receiver was replaced with a human. The results showed that 68\% of the cases, human participants picked the correct image.

Similarly, Mordatch and Abbeel~\cite{mordatch2017emergence} investigated the emergence of language with the difference that in their setting there were no explicit roles for the agents (i.e., sender or receiver). To learn, they proposed an end-to-end differentiable model of all agent and environment state dynamics over time to calculate the gradient of the return with backpropagation.

\subsection{Learning communication}
\label{sec:learningCommunication}

As we discussed in the previous section, one of the desired emergent behaviors of multiagent interaction is the emergence of communication~\cite{Lazaridou:2016un,mordatch2017emergence}. 
This setting usually considers a set of \emph{cooperative agents} in a \emph{partially observable} environment (see Section~\ref{sec:deeprl}) where agents need to maximize their shared utility by means of communicating information.

Reinforced Inter-Agent Learning (RIAL) and Differentiable Inter-Agent Learning (DIAL) are two methods using deep networks to learn to communicate~\cite{Foerster:2016ud}. Both methods use a neural net that outputs the agent's $Q$ values (as done in standard DRL algorithms) and a message to communicate to other agents in the next timestep. RIAL is based on DRQN and also uses the concept of \emph{parameter sharing}, i.e., using a single network whose parameters are shared among all agents. In contrast, DIAL directly passes gradients via the communication channel during learning, and  messages are discretized and mapped to the set of communication actions during execution. 

\emph{Memory-driven} (MD) communication was proposed on top of the Multi-Agent Deep Deterministic Policy Gradient  (MADDPG)~\cite{Lowe:2017wb} method. In MD-MADDPG \cite{pesce2019improving}, the agents use a shared memory as a communication channel: before taking an action, the agent first reads the memory, then writes a response. In this case the agent's policy becomes dependent on its private observation and its interpretation of the collective memory. Experiments were performed with two agents in cooperative scenarios. The results highlighted the fact that the communication channel was used differently in each environment, e.g., in simpler tasks agents significantly decrease their memory activity near the end of the task as there are no more changes in the environment; in more complex environments, the changes in memory usage appear at a much higher frequency due to the presence of many sub-tasks.

Dropout~\cite{srivastava2014dropout} is a technique to prevent overfitting (in supervised learning this happens when the learning algorithm achieves good performance only on a specific data set and fails to generalize) in neural networks which is based on randomly dropping units and their connections during training time. Inspired by dropout, Kim et al.~\cite{kim2019message} proposed a similar approach in multiagent environments where direct communication through messages is allowed. In this case, the messages of other agents are dropped out at training time, thus the authors proposed the Message-Dropout MADDPG algorithm~\cite{kim2019message}. This method is expected to work in fully or limited communication environments. The empirical results show that with properly chosen message dropout rate, the proposed method both significantly improves the training speed and the robustness of learned policies (by introducing communication errors) during execution time. This capability is important as MDRL agents trained in simulated or controlled environments will be less fragile when transferred to more realistic environments.

While RIAL and DIAL used a discrete communication channel, CommNet~\cite{Sukhbaatar:2016tk} \revise{used} a continuous vector channel. Through this channel agents receive the summed transmissions of other agents. The authors assume full cooperation and train a single network for all the agents. There are two distinctive characteristics of CommNet from previous works: it allows multiple communication cycles at each timestep and a dynamic variation of agents at run time, i.e., agents come and go in the environment. %

In contrast to previous approaches, in Multiagent Bidirectionally Coordinated Network (BiCNet)~\cite{Peng:vd}, communication takes place in the latent space (i.e., in the hidden layers). It also uses parameter sharing, however, it proposes bidirectional  recurrent neural networks~\cite{Schuster:1997ur} to model the actor and critic networks of their model. Note that in BiCNet agents do not \emph{explicitly} share a message and thus it can be considered a method for learning cooperation.

Learning communication is an active area in \MDRL with many open questions, in this context, we refer the interested reader to a recent work by Lowe et al.~\cite{lowe2019pitfalls} where it discusses common pitfalls (and recommendations to avoid those) while measuring communication in multiagent environments.

\subsection{Learning cooperation}
\label{sec:learningCooperation}

Although \emph{explicit communication} is a new emerging trend in \MDRL, there has already been a large amount of work in MAL for cooperative settings\footnote{There is \revise{a} large body of research on coordinating multiagent teams by specifying communication protocols~\cite{tambe1997towards,grosz1996collaborative}\revise{:} these expect agents to know the team's goal as well as the tasks required to accomplish the goal.} that do not involve communication~\cite{Panait:2005wj,Matignon:2012bj}. Therefore, it was a natural starting point for many recent \MDRL works.

Foerster et al.~\cite{Foerster:2017ti} studied the simple scenario of \emph{cooperation with independent Q-learning agents} (see Section~\ref{sec:multiagentlearning}), where the agents use the standard DQN architecture of neural networks and an experience replay buffer (see Figure~\ref{fig:dqn_diagram}). However, for the ER to work, the data distribution needs to follow certain assumptions (see Section~\ref{sec:deeprl}) which are no loger valid due to the multiagent nature of the world: the dynamics that generated the data in the ER no longer reflect the current dynamics, making the experience obsolete~\cite{Foerster:2017ti,lin1992self}. Their solution is to add information to the experience tuple that can help to \emph{disambiguate the age of the sampled data} from the replay memory. Two approaches were proposed. The first is Multiagent Importance Sampling which adds the probability of the joint action so an importance sampling correction~\cite{bishop2006pattern,precup2000eligibility} can computed when the tuple is later sampled for training. This was similar to previous works in adaptive importance sampling~\cite{frank2008reinforcement,ahamed2006adaptive} and off-environment RL~\cite{ciosek2017offer}. The second approach is Multiagent Fingerprints which adds the estimate (i.e., fingerprint) of other agents' policies (loosely inspired by Hyper-Q~\cite{Tesauro:2003wq}, see Section~\ref{sec:bridging:examples}). For the practical implementation, good results were obtained by using the training iteration number and exploration rate as the fingerprint.

Gupta et al.~\cite{10.1007/978-3-319-71682-4_5} tackled cooperative environments in partially observable domains without explicit communication. They proposed  \emph{parameter sharing} (PS) as a way to improve learning in homogeneous multiagent environments (where agents have the same set of actions). The idea is to have one globally shared learning network that can still behave differently in execution time, i.e., because its inputs (individual agent observation and agent index) will be different. %
They tested three variations of this approach with parameter sharing: PS-DQN, PS-DDPG and PS-TRPO, which extended single-agent DQN, DDPG and TRPO algorithms, respectively. The results showed that PS-TRPO outperformed the other two. Note that Foerster et al.~\cite{Foerster:2016ud} concurrently proposed a similar concept, see Section~\ref{sec:learningCommunication}. %

Lenient-DQN (LDQN)~\cite{Palmer:2018wv} took the \emph{leniency} concept~\cite{Bloembergen:2010wd} (originally presented in MAL) and extended their use to \MDRL. \revise{The purpose of leniency is to overcome a pathology called relative overgeneralization~\cite{Wei:2016wq,Panait:2006tx,panait2008theoretical}. Similar to other approaches designed to overcome relative overgeneralization (e.g., distributed Q-learning~\cite{lauer2000algorithm} and hysteretic Q-learning~\cite{Matignon:2012bj}) lenient learners initially maintain an optimistic disposition to mitigate the noise from transitions resulting in miscoordination, preventing agents from being drawn towards sub-optimal but wide peaks in the reward search space~\cite{palmer2018negative}.} %
However, similar to other MDRL works~\cite{Foerster:2017ti}, the LDQN authors experienced problems with the ER buffer and arrived at a similar solution: adding information to the experience tuple, in their case, the leniency value. When sampling from the ER buffer, this value is used to determine a leniency condition; if the condition is not met then the sample is ignored.

In a similar vein, Decentralized-Hysteretic Deep Recurrent Q-Networks (DEC-HDRQNs)~\cite{Omidshafiei:2017uk} were proposed for fostering cooperation among independent learners. The motivation is similar to LDQN, making an optimistic value update, however, their solution is different. Here, the authors took inspiration from  Hysteretic Q-learning~\cite{Matignon:2012bj}, originally presented in MAL, where two learning rates were used. \revise{A difference between lenient agents and hysteretic Q-learning is that lenient agents are only \emph{initially} forgiving towards teammates. Lenient learners over time apply less leniency towards updates that would lower utility values, taking into account how frequently observation-action pairs have been encountered. The idea being that the transition from optimistic to average reward learner will help make lenient learners more robust towards misleading stochastic rewards~\cite{Bloembergen:2010wd}.} Additionally, in DEC-HDRQNs the ER buffer is also extended into \emph{concurrent experience replay trajectories}, which are composed of three dimensions: agent index, the episode, and the timestep; when training, the sampled traces have the same starting timesteps. Moreover, to improve on generalization over different tasks, i.e., multi-task learning\cite{caruana1997multitask}, DEC-HDRQNs make use of policy distillation~\cite{Rusu:2015td,hinton2015distilling} (see Section~\ref{sec:bridging:examples}). In contrast to other approaches, DEC-HDRQNS are fully decentralized during learning and execution.

Weighted Double Deep Q-Network (WDDQN)~\cite{zheng2018weighted} is based on having double estimators. This idea was originally introduced in Double Q-learning~\cite{hasselt2010double} and aims to remove the existing overestimation bias caused by using the maximum action value as an approximation for the maximum expected action value (see Section~\ref{sec:bridging:examples}). It also uses a \emph{lenient} reward~\cite{Bloembergen:2010wd} to be optimistic during initial phase of coordination and proposes a \emph{scheduled} replay strategy in which samples closer to the terminal states are heuristically given higher priority;  this strategy might not be applicable for any domain. For other works extending the ER to multiagent settings see MADDPG~\cite{Lowe:2017wb}, Sections~\ref{sec:bridging:examples} and \ref{sec:bridging:lessons}.

\begin{figure}
\centering
\includegraphics[width=0.99\textwidth]{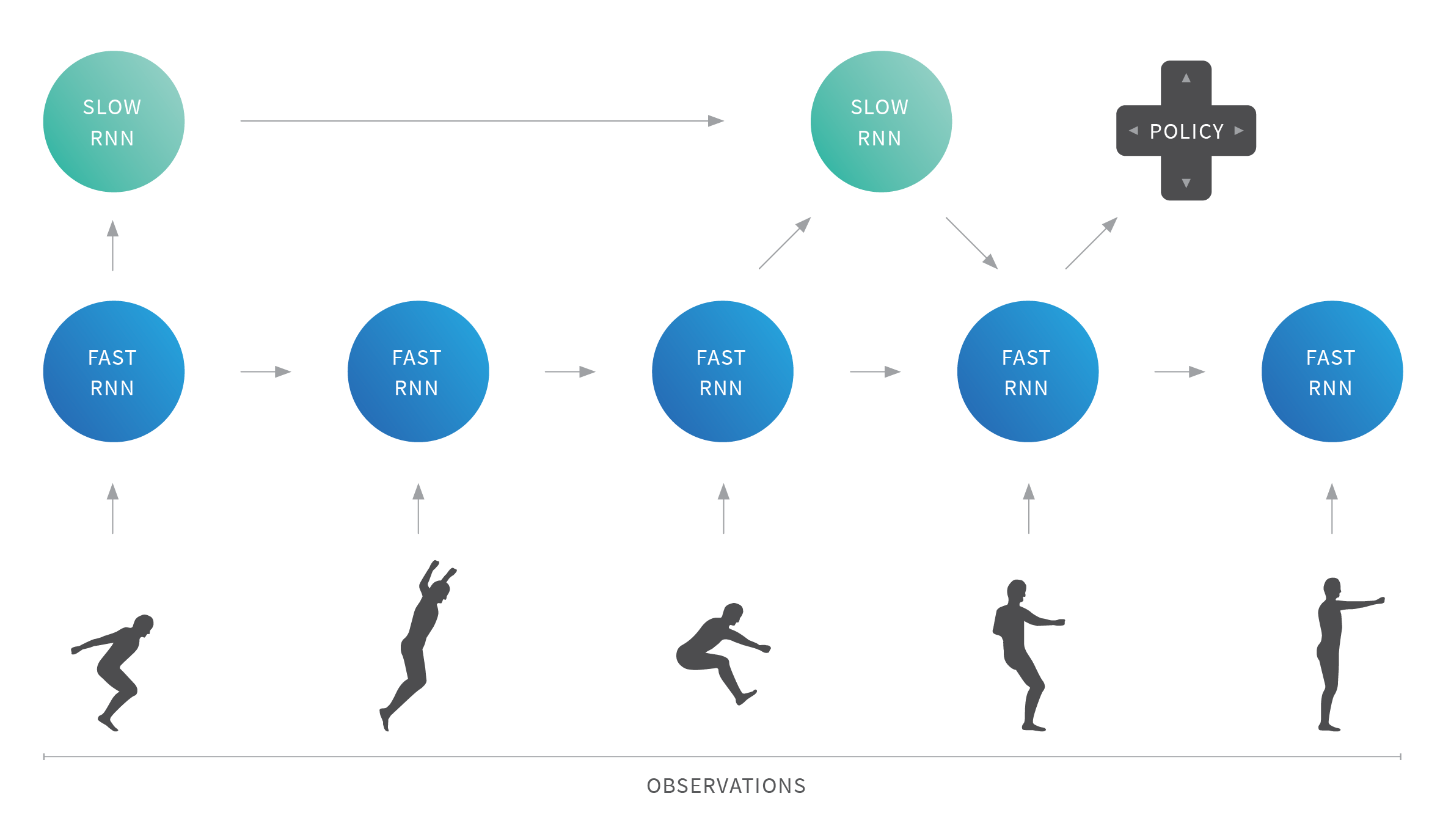}
\caption{A schematic view of the architecture used in FTW (For the Win)~\cite{Jaderberg:2018wx}: two unrolled recurrent neural networks (RNNs) operate at different time-scales, the idea is that the \emph{Slow RNN} helps with long term temporal correlations. Observations are latent space output of some convolutional neural network to learn non-linear features. Feudal Networks~\cite{Vezhnevets:2017wl} is another work in single-agent DRL that also maintains a multi-time scale hierarchy where the slower network sets the goal, and the faster network tries to achieve them. Fedual Networks were in turn, inspired by early work in RL which proposed a hierarchy of Q-learners~\cite{dayan1993feudal,singh1992transfer}.
}
\label{fig:ftw}
\end{figure}

While previous approaches were mostly inspired by how MAL algorithms could be extended to \MDRL, other works take as base the results by single-agent DRL.
One example is the For The Win (FTW)~\cite{Jaderberg:2018wx} agent which is based on the \revise{actor-learner structure of} IMPALA~\cite{espeholt2018impala} (see Section~\ref{sec:deeprl}). \revise{The authors test FTW in} a game where two opposing teams compete to capture each other's flags~\cite{capturetheflag}. To deal with the MAL problem they propose two main additions: a \emph{hierarchical two-level representation} with recurrent neural networks  operating at different timescales, as depicted in Figure~\ref{fig:ftw}, and a \emph{population based training}~\cite{rosin1997new,lehman2008exploiting,jaderberg2017population} where 30 agents were trained in parallel together with a stochastic matchmaking scheme that biases agents to be of similar \emph{skills}. The Elo rating system~\cite{elo1978rating} was originally devised to rate chess player skills,\footnote{Elo uses a normal distribution for each player skill, and after each match, both players' distributions are updated based on measure of \emph{surprise}, i.e., if a user with previously lower (predicted) skill beats a high skilled one, the low-skilled player is significantly increased.} TrueSkill~\cite{herbrich2007trueskill} extended Elo by tracking uncertainty in skill rating, supporting draws, and matches beyond 1 vs 1\revise{; $\alpha-$Rank is a more recent alternative to ELO~\cite{omidshafiei2019alpha}}. FTW did not use TrueSkill but a simpler extension of Elo for $n$ vs $n$ games (by adding individual agent ratings to compute the team skill). Hierarchical approaches were previously proposed in RL, e.g., Feudal RL~\cite{dayan1993feudal,singh1992transfer}, \revise{and were} later extended to DRL in Feudal networks~\cite{Vezhnevets:2017wl}; population based training can be considered analogous to evolutionary strategies that employ self-adaptive hyperparameter tuning to modify how the genetic algorithm itself operates~\cite{lehman2008exploiting,back1996evolutionary,de2006evolutionary}. An interesting result from FTW is that the population-based training obtained better results than training via self-play~\cite{tesauro1995temporal}, which was a standard concept in previous works~\cite{Silver:2016hl,Bowling:2015ub}.
FTW used heavy compute resources, it used 30 agents (processes) in parallel where every training game lasted 4500 agent steps ($\approx$ five minutes) and agents were trained for two billion steps ($\approx$ 450K games).

Lowe et al.~\cite{Lowe:2017wb} noted that using standard policy gradient methods (see Section~\ref{sec:rl}) on multiagent environments yields high variance and performs poorly. 
This occurs because the variance is further increased as all the agents' rewards depend on the rest of the agents, and it is formally shown that as the number of agents increase, the probability of taking a correct gradient direction decreases exponentially~\cite{Lowe:2017wb}. Therefore, to overcome this issue Lowe et al. proposed the
Multi-Agent Deep Deterministic Policy Gradient (MADDPG)~\cite{Lowe:2017wb}, building on DDPG~\cite{lillicrap2015continuous} (see Section~\ref{sec:deeprl}), to train a centralized critic per agent that is given all agents' policies during training to reduce the variance by removing the non-stationarity caused by the concurrently learning agents. %
Here, the actor only has local information (turning the method into a centralized training with decentralized execution) and the ER buffer records experiences of \emph{all} agents. MADDPG was tested in both cooperative and competitive scenarios, experimental results show that it performs better than several decentralized methods (such as DQN, DDPG, and TRPO). The authors mention that traditional RL methods do not produce consistent gradient signals. This is exemplified in a challenging competitive scenarios where agents continuously adapt to each other causing the learned best-response policies oscillate --- for such a domain, MADDPG is shown to learn more robustly than DDPG.

Another approach based on policy gradients is the Counterfactual Multi-Agent Policy Gradients (COMA)~\cite{Foerster:2017uq}. COMA \revise{was designed for} the fully centralized setting and the \emph{multiagent credit assignment problem}~\cite{Tumer:2007vs}, i.e., how the agents should deduce their contributions when learning in a cooperative setting in the presence of only global rewards. Their proposal is to compute a \emph{counterfactual baseline}, that is, marginalize out the action of the agent while keeping the rest of the other agents' actions fixed. Then, an advantage function can be computed comparing the current $Q$ value to the counterfactual. This counterfactual baseline has its roots in \emph{difference rewards}, which is a method for obtaining the individual contribution of an agent in a cooperative multiagent team~\cite{Tumer:2007vs}. In particular, the \emph{aristocrat} utility aims to measure the difference between an agent's actual action and the average action~\cite{wolpert2002optimal}. The intention would be equivalent to sideline the agent by having the agent perform an action where the reward does not depend on the agent's actions, i.e., to consider the reward that would have arisen assuming a world without that agent having ever existed (see Section~\ref{sec:bridging:lessons}).

On the one hand, fully centralized approaches (e.g., COMA) do not suffer from non-stationarity but have constrained scalability. On the other hand, independent learning agents are better suited to scale but suffer from non-stationarity issues. There are some hybrid approaches that learn \emph{a centralized but factored} $Q$ value function~\cite{guestrin2002multiagent,kok2004sparse}. Value Decomposition Networks (VDNs)~\cite{Sunehag:2018vua} decompose a team value function into an additive decomposition of the individual value functions. Similarly, QMIX~\cite{Rashid:2018wy} relies on the idea of factorizing, however, instead of sum, QMIX assumes a \emph{mixing network} that combines the local values in a non-linear way, which can represent monotonic action-value functions. While the mentioned approaches have obtained good empirical results, the factorization of value-functions in multiagent scenarios using function approximators (\MDRL) is an ongoing research topic, with open questions such as how well factorizations capture complex coordination problems and how to learn those factorizations~\cite{castellini2019representational} (see Section~\ref{sec:challengesmdrl}).

\subsection{Agents modeling agents}
\label{sec:agentsModeling}

\begin{figure}[t]
\centering
\includegraphics[scale=0.55]{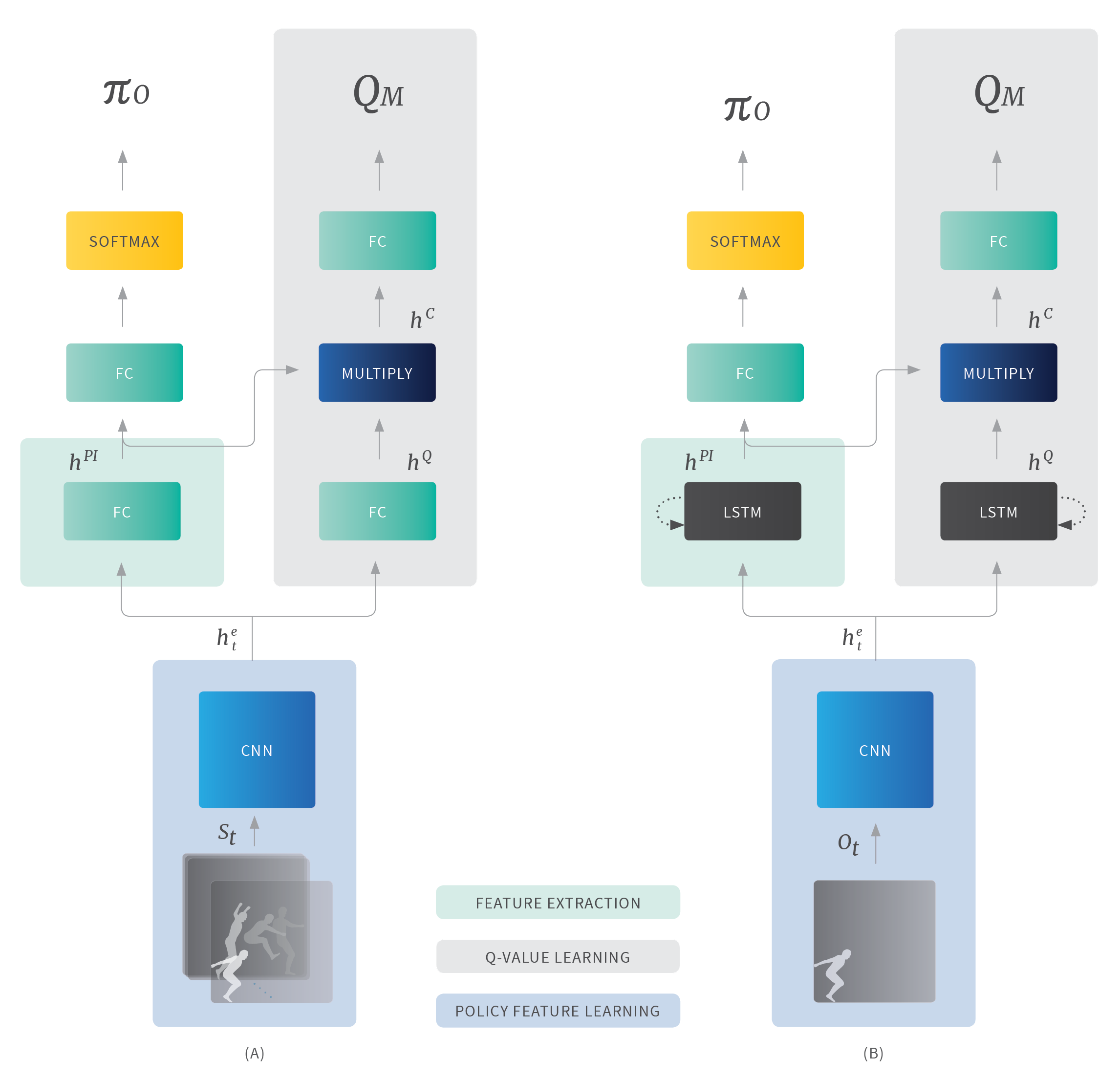}
\caption{(a) Deep Policy Inference Q-Network: receives four stacked frames as input (similar to DQN, see Figure~\ref{fig:dqn_filter}). (b) Deep Policy Inference Recurrent Q-Network: receives one frame as input and has an LSTM layer instead of a fully connected layer (FC). Both approaches~\cite{Hong:2018vp} condition the $Q_M$ value outputs on the policy features, $h^{PI}$, which are also used to learn the opponent policy $\pi_o$.}
\label{fig:DPIRQN}
\end{figure}

An important ability for agents to have is to reason about the behaviors of other agents by constructing models that make predictions about the modeled agents~\cite{Albrecht:2018dp}. An early work for modeling agents while using deep neural networks was the
Deep Reinforcement Opponent Network (DRON)~\cite{He:2016up}. The idea is to have two networks: one which evaluates $Q$-values and a second one that \emph{learns a representation of the opponent's policy}. Moreover, the authors proposed to have several expert networks to combine their predictions to get the estimated $Q$ value, the idea being that each expert network captures one type of opponent strategy~\cite{Gmytrasiewicz:2005un}. This is related to previous works in type-based reasoning from game theory~\cite{harsanyi1967games,kalai1993rational} later applied in AI~\cite{Gmytrasiewicz:2005un,Albrecht:2018dp,Barrett:2013uh}. The mixture of experts idea was presented in supervised learning where each expert handled a subset of the data (a subtask), and then a gating network decided which of the experts should be used~\cite{jacobs1991adaptive}.

DRON uses hand-crafted features to define the opponent network. In contrast, Deep Policy Inference Q-Network (DPIQN) and its recurrent version, DPIRQN~\cite{Hong:2018vp} learn \emph{policy features} directly from raw observations of the other agents. The way to learn these policy features is by means of auxiliary tasks~\cite{Jaderberg:2016uv,sutton2011horde} (see Sections~\ref{sec:deeprl} and \ref{sec:bridging:examples}) that provide additional learning goals, in this case, the auxiliary task is to learn the opponents' policies. This auxiliary task modifies the loss function by computing an auxiliary loss: the cross entropy loss between the inferred opponent policy and the ground truth (one-hot action vector) of the opponent. Then, the $Q$ value function of the learning agent is conditioned on the opponent's policy features (see Figure~\ref{fig:DPIRQN}), which aims to reduce the non-stationarity of the environment. The authors used an adaptive training procedure to adjust the attention (a weight on the loss function) to either emphasize learning the policy features (of the opponent) or the respective $Q$ values of the agent. An  advantage of these approaches is that modeling the agents can work for both opponents and teammates~\cite{Hong:2018vp}.

In many previous works an opponent model is learned from observations. Self Other Modeling (SOM)~\cite{Raileanu:2018tb} proposed a different approach, this is, using \emph{the agent's own policy as a means to predict the opponent's actions}. SOM can be used in cooperative and competitive settings (with an arbitrary number of agents) and infers other agents' goals. This is important because in the evaluated domains, the reward function depends on the goal of the agents. SOM uses two networks, one used for computing the agents' own policy, and a second one used to infer the opponent's goal. The idea is that these networks have the same input parameters but with different values (the agent's or the opponent's). In contrast to previous approaches, SOM is not focused on learning the opponent policy, i.e., a probability distribution over next actions, but rather on estimating the opponent's goal. SOM is expected to work best when agents share a set of goals from which each agent gets assigned one at the beginning of the episode and the reward structure depends on both of their assigned goals. Despite its simplicity, training takes longer as an additional optimization step is performed given the other agent's observed actions.

There is a long-standing history of combining game theory and MAL~\cite{Shoham:2007vw,nowe2012game,Bowling:2015ub}. From that context, some approaches were inspired by influential game theory approaches. Neural Fictitious Self-Play (NFSP)~\cite{Heinrich:69lSVXjW} builds on fictitious (self-) play~\cite{Brown:1951vc,heinrich2015fictitious}, together with two deep networks to find \emph{approximate Nash equilibria}\footnote{Nash equilibrium~\cite{Nash:1950vo} is a solution concept in game theory in which no agent would choose to deviate from its strategy (they are a best response to others' strategies). This concept has been explored in seminal MAL algorithms like Nash-Q learning~\cite{Hu:2003wb} and Minimax-Q learning~\cite{Littman:1994ta,Littman:2001vc}.}
in two-player imperfect information games~\cite{VonNeumann:1945tc} (for example, consider Poker: when it is an agent's turn to move it does not have access to all information about the world).
One network learns an \emph{approximate best response} ($\epsilon-$greedy over $Q$ values) to the historical behavior of other agents and the second one (called the average network) learns to imitate its own past best response behaviour  using supervised classification. The agent behaves using a mixture of the average and the best response networks depending on the probability of an anticipatory parameter~\cite{shamma2005dynamic}. Comparisons with DQN in Leduc Hold’em Poker revealed that DQN's deterministic strategy is highly exploitable. Such strategies are sufficient to behave optimally in single-agent domains, i.e., MDPs for which DQN was designed. However, imperfect-information games generally require stochastic strategies to achieve optimal behaviour~\cite{Heinrich:69lSVXjW}. DQN learning experiences are both highly correlated over time, and highly focused on a narrow state distribution. In contrast to NFSP agents whose experience varies more smoothly, resulting in a more stable data distribution, more stable neural networks and better performance.

The (N)FSP concept was further generalized in Policy-Space Response Oracles (PSRO)~\cite{Lanctot:vn}, where it was shown that fictitious play is one specific meta-strategy distribution over a set of previous (approximate) best responses (summarized by a meta-game obtained by empirical game theoretic analysis~\cite{Walsh:2002ur}), but there are a wide variety to choose from. One reason to use mixed meta-strategies is that it prevents overfitting\footnote{Johanson et al.~\cite{johanson2011accelerating} also found ``overfitting"  when solving large extensive games (e.g., poker) --- the performance in an abstract game improved but it was worse in the full game.} the responses to one specific policy, %
and hence provides a form of opponent/teammate regularization. \revise{An approximate scalable version} of the algorithm leads to a graph of agents best-responding independently called Deep Cognitive Hierarchies (DCHs)~\cite{Lanctot:vn} due to its similarity to behavioral game-theoretic models~\cite{Camerer:2004va,CostaGomes:2001wt}.

Minimax is a paramount concept in game theory that is roughly described as minimizing the worst case scenario (maximum loss)~\cite{VonNeumann:1945tc}. Li et al.~\cite{li2019robust} took the minimax idea as an approach to robustify learning in multiagent environments so that the learned robust policy should be able to behave well even with strategies not seen during training. They extended the MADDPG algorithm~\cite{Lowe:2017wb} to Minimax Multiagent Deep Deterministic Policy Gradients (M3DDPG), which updates policies considering a worst-case scenario: assuming that all other agents act adversarially. This yields a minimax learning objective which is computationally intractable to directly optimize. They address this issue by taking ideas from robust reinforcement learning~\cite{morimoto2005robust} which implicitly adopts the minimax idea by using the \emph{worst noise} concept~\cite{pinto2017robust}. In MAL different approaches were proposed to assess the robustness of an algorithm, e.g., guarantees of safety~\cite{Chakraborty:2013ii,Powers:2007gq}, security~\cite{Crandall:2011dt} or  exploitability~\cite{Johanson:2007ts,McCracken:2004ua,Damer:2017us}.

Previous approaches usually learned a model of the other agents as a way to predict their behavior. However, they do not explicitly \emph{account for anticipated learning of the other agents}, which is the objective of Learning with Opponent-Learning Awareness (LOLA)~\cite{Foerster:2018to}. LOLA optimizes the expected return \emph{after the opponent updates its policy one step}.
Therefore, a LOLA agent directly shapes the policy updates of other agents to maximize its own reward. One of LOLA's assumptions is having access to opponents' policy parameters.
LOLA builds on previous ideas by Zhang and Lesser~\cite{zhang2010multi} where the learning agent predicts the opponent's policy parameter update but only uses it to learn a best response (to the anticipated updated parameters).

Theory of mind is part of a group of \emph{recursive reasoning} approaches\cite{Gmytrasiewicz:2000tx,Gmytrasiewicz:2005un,Camerer:2004vt,carmel1996incorporating} in which agents have explicit beliefs about the mental states of other agents. The mental states of other agents may, in turn, also contain beliefs and mental states of other agents, leading to a nesting of beliefs~\cite{Albrecht:2018dp}.
Theory of Mind Network (ToMnet)~\cite{Rabinowitz:2018uf} starts with a simple premise: when encountering a novel opponent, \emph{the agent should already have a strong and rich prior about how the opponent should behave}. \revise{ToMnet has} an architecture composed of three networks: (i) a character network that learns from historical information, (ii) a mental state network that takes the character output and the recent trajectory, and (iii) the prediction network that takes the current state as well as the outputs of the other networks as its input. The output of the architecture is open for different problems but in general its goal is to predict the opponent's next action. A main advantage of ToMnet is that it can predict general behavior, for all agents; or specific, for a particular agent.}

Deep Bayesian Theory of Mind Policy (Bayes-ToMoP)~\cite{Yang:4pJjEXeO} is another algorithm that takes inspiration \revise{from} theory of mind~\cite{deWeerd:2013cg}. \revise{The algorithm} assumes the opponent has different stationary strategies to act and changes among them over time~\cite{HernandezLeal:2017iy}. Earlier work in MAL dealt with this setting, e.g., BPR+~\cite{HernandezLeal:2016tw} extends the Bayesian policy reuse\footnote{Bayesian policy reuse assumes an agent with prior experience in the form of a library of policies. When a novel task instance occurs, the objective is to reuse a policy from its library based on observed signals which correlate to policy performance~\cite{Rosman:2015vh}.} framework~\cite{Rosman:2015vh} to multiagent settings \revise{(BPR assumes a single-agent environment; BPR+ aims to best respond to the opponent in a multiagent game)}. A limitation of BPR+ is that it behaves poorly against itself (self-play), thus, Deep Bayes-ToMoP uses theory of mind to provide a higher-level reasoning strategy which provides an optimal behavior against BPR+ agents. %

Deep BPR+~\cite{zheng2018deep} is another work inspired by BPR+ which uses neural networks as value-function approximators. It not only uses the environment reward but also uses the online learned opponent model~\cite{HernandezLeal:2016twa,HernandezLeal:2017vi} to construct a rectified belief over the opponent strategy. Additionally, it \revise{leverages} ideas from policy distillation~\cite{Rusu:2015td,hinton2015distilling} and extends them to the multiagent case to create a distilled policy network. In this case, whenever a new acting policy is learned, distillation is applied to consolidate the new updated library which improves in terms of storage and generalization (over opponents).

\section{Bridging RL, MAL and \MDRL}
\label{sec:Bridging}

This section aims to provide directions to promote fruitful cooperations between sub-communities. First, we address the pitfall of \emph{deep learning amnesia}, roughly described as missing citations to the original works and not exploiting the advancements that have been made in the past. We present examples on how ideas originated earlier, for example in RL and MAL, were successfully extended to \MDRL (see Section~\ref{sec:bridging:examples}). Second, we outline \emph{lessons learned} from the works analyzed in this survey (see Section~\ref{sec:bridging:lessons}). Then we point the readers to recent benchmarks for MDRL (see Section~\ref{sec:benchmarksmdrl}) and we discuss the practical challenges that arise in MDRL like high computational demands and reproducibility (see Section~\ref{sec:challengesmdrl}). Lastly, we pose some open research challenges and reflect on their relation with previous open questions in MAL~\cite{Albrecht:2018dp} (see Section~\ref{sec:bridging:questions}).

\subsection{Avoiding deep learning amnesia: examples in \MDRL}
\label{sec:bridging:examples}

This survey focuses on recent \emph{deep} works, however, in previous sections, when describing recent algorithms, we also point to original works that inspired them. Schmidhuber said ``Machine learning is the science of credit assignment. The machine learning community itself profits from proper credit assignment to its members"~\cite{schmidhuber2015critique}. In this context, we want to avoid committing the pitfall of not giving credit to original ideas that were proposed earlier, 
a.k.a. \emph{deep learning amnesia}. Here, we provide some specific examples of research milestones that were studied earlier, e.g., RL or MAL, and that now became highly relevant for MDRL. Our purpose is to highlight that existent literature contains pertinent ideas and algorithms that should not be ignored. On the contrary, they should be examined and cited~\cite{cite2017Arxiv,collaboration2019trust} to understand recent developments~\cite{origin2017deeplearning}.

\paragraph{Dealing with non-stationarity in independent learners}
It is well known that using independent learners makes the environment non-stationary from the agent's point of view~\cite{Tuyls:2012up,laurent2011world}. Many MAL algorithms tried to solve this problem in different ways~\cite{HernandezLeal:2017wd}. One example is \emph{Hyper-Q}~\cite{Tesauro:2003wq} which accounts for the (values of mixed) strategies of other agents and includes that information in the state representation, which effectively turns the learning problem into a stationary one.  Note that in this way it is possible to even consider adaptive agents. 
Foerster et al.~\cite{Foerster:2016ud} make use of this insight to propose their \emph{fingerprint} algorithm in an \MDRL problem (see Section~\ref{sec:learningCooperation}). Other examples include the leniency concept~\cite{Bloembergen:2010wd} and Hysteretic Q-learning~\cite{Matignon:2012bj} originally presented in MAL, which now have their ``deep" counterparts, LDQNs~\cite{Palmer:2018wv} and DEC-HDRQNs\cite{Omidshafiei:2017uk}, see Section~\ref{sec:learningCooperation}.

\paragraph{Multiagent credit assignment}
In cooperative multiagent scenarios, it is common to use either \emph{local rewards}, unique for each agent, or \emph{global rewards}, which represent the entire group's performance~\cite{Agogino:2008wh}. However, local rewards are usually harder to obtain, therefore, it is common to rely only on the global ones. This raises the problem of \emph{credit assignment}: how do\revise{es} a single agent's actions contribute to a system that involves the actions of many agents~\cite{Agogino:2004ug}. A solution that came from MAL research that has proven successful in many scenarios is \emph{difference rewards}~\cite{Tumer:2007vs,Agogino:2008wh,Devlin:2014vf}, which aims to capture an agent's
contribution to the system's global performance. In particular the \emph{aristocrat} utility aims to measure the difference between an agent’s actual action and the average action~\cite{wolpert2002optimal}, however, it has a self-consistency problem and in practice it is more common to compute the \emph{wonderful life utility}~\cite{wolpert1999general,wolpert2002optimal}, which proposes to use a clamping operation that would be equivalent to removing that player from the team. COMA~\cite{Foerster:2017uq} builds on these concepts to propose an \emph{advantage function} based on the contribution of the agent, which can be efficiently computed with deep neural networks (see Section~\ref{sec:learningCooperation}).

\paragraph{Multitask learning}
In the context of RL, multitask learning~\cite{caruana1997multitask} is an area that develops agents that can act in \emph{several related tasks} rather than just in a single one~\cite{Taylor:2009ur}. \emph{Distillation}, roughly defined as transferring the knowledge from a large model to a small model, was a concept originally introduced for supervised learning and model compression~\cite{bucilu2006model,hinton2015distilling}. Inspired by those works, Policy distillation~\cite{Rusu:2015td} was extended to the DRL realm. Policy distillation was used to train a much smaller network and to merge \emph{several task-specific policies} into a single policy, i.e., for multitask learning. In the \MDRL setting, Omidshafiei et al.~\cite{Omidshafiei:2017uk} successfully adapted policy distillation within Dec-HDRQNs to obtain a more general multitask multiagent network (see Section~\ref{sec:learningCooperation}). Another example is Deep BPR+~\cite{zheng2018deep} which uses distillation to generalize over multiple opponents (see Section~\ref{sec:agentsModeling}).

\paragraph{Auxiliary tasks}
Jaderberg et al.~\cite{Jaderberg:2016uv} introduced the term auxiliary task with the insight that (single-agent) environments contain a variety of possible training signals (e.g., pixel changes). These tasks are naturally implemented in DRL in which the last layer is split into multiple parts (heads), each working on a different task. All heads propagate errors into the same shared preceding part of the network, which would then try to form representations, in its next-to-last layer, to support all the heads~\cite{sutton2018reinforcement}. However, the idea of multiple predictions about arbitrary signals was originally suggested for RL, in the context of general value functions~\cite{sutton2011horde,sutton2018reinforcement} and there still open problems, for example, better theoretical understanding~\cite{bellemare2019geometric,du2018adapting}. In the context of neural networks, early work proposed \emph{hints} that improved the network performance and learning time. Suddarth and Kergosien~\cite{suddarth1990rule} presented a minimal example of a small neural network where it was shown that adding an auxiliary task effectively removed local minima. One could think of extending these auxiliary tasks to modeling other agents' behaviors~\cite{hern2019agent,mordatch2017emergence}, which is one of the key ideas that DPIQN and DRPIQN~\cite{Hong:2018vp} proposed in \MDRL settings (see Section~\ref{sec:agentsModeling}). 

\paragraph{Experience replay} 
Lin~\cite{lin1992self,lin1991programming} proposed the concept of experience replay to speed
up the credit assignment propagation process in single agent RL. This concept became central to many DRL works~\cite{Mnih:2013wp} (see Section~\ref{sec:deeprl}). However, Lin stated that a condition for the ER to be useful is that ``the environment should not change over time because this makes past experiences irrelevant or even harmful"~\cite{lin1992self}. This is a problem in domains where many agents are learning since the environment becomes non-stationary from the point of view \revise{of} each agent. Since DRL relies heavily on experience replay, this is an issue in MDRL: the non-stationarity introduced means that the dynamics that generated the data in the agent's replay memory no longer reflect the current dynamics in which it is learning~\cite{Foerster:2016ud}. To overcome this problem different methods have been proposed~\cite{Foerster:2017ti,Palmer:2018wv,Omidshafiei:2017uk,zheng2018weighted}, see Section~\ref{sec:bridging:lessons}.

\paragraph{Double estimators}
Double Q-learning~\cite{hasselt2010double} proposed to reduce the overestimation of action values in Q-learning, this is caused \revise{by} using the maximum action value as an approximation for the maximum expected action value. Double Q-learning works by keeping two Q functions and was proven to convergence to the optimal policy~\cite{hasselt2010double}. Later this idea was applied to arbitrary function approximators, including deep neural networks, i.e., Double DQN~\cite{van2016deep}, which were naturally applied since two networks were already used in DQN (see Section \ref{sec:deeprl}). These ideas have also been recently applied to MDRL~\cite{zheng2018weighted}.

\subsection{Lessons learned}
\label{sec:bridging:lessons}

We have exemplified how RL and MAL can be extended for \MDRL settings. Now, we outline general \emph{best practices} learned from the works analyzed throughout this paper.

\begin{itemize}
\item \textit{Experience replay buffer in \MDRL.}
While some works removed the ER buffer in \MDRL~\cite{Foerster:2016ud} it is an important component in many DRL and \MDRL algorithms. However, using the standard buffer (i.e., keeping $\langle s,a,r,s' \rangle$) will probably fail due to a lack of theoretical guarantees under this setting, see Sections~\ref{sec:deeprl} and \ref{sec:bridging:examples}. \emph{Adding information in the experience tuple} that can help disambiguate the sample is the solution adopted in many works, whether a value based method~\cite{Foerster:2017ti,Palmer:2018wv,Omidshafiei:2017uk,zheng2018weighted} or a policy gradient method~\cite{Lowe:2017wb}. In this regard, it is an open question to consider \revise{how} new DRL ideas \revise{could be best integrated into} the ER~\cite{Andrychowicz:2017vq,Schaul:2015ty,Lipton:rDKEj8a8,de2018experience,isele2018selective} and how those \revise{ideas} would fare in a \MDRL setting.

\item \textit{Centralized learning with decentralized execution.} 
Many MAL works were either fully centralized or fully decentralized approaches. However, inspired by \emph{decentralized partially observable
Markov decison processes} (DEC-POMDPs)~\cite{bernstein2002complexity,oliehoek2016concise},\footnote{Centralized planning and decentralized execution is also a standard paradigm for multiagent planning~\cite{oliehoek2008optimal}.} in \MDRL this new mixed paradigm has been commonly used ~\cite{Foerster:2017ti,Palmer:2018wv,Rashid:2018wy,Lanctot:vn,Foerster:2017uq,Lowe:2017wb} (a notable exception are DEC-HDRQNs~\cite{Omidshafiei:2017uk} which perform learning and execution in a decentralized manner, see Section~\ref{sec:learningCooperation}). Note that not all real-world problems fit into this paradigm and it is more common for robotics or games where a simulator is generally available~\cite{Foerster:2016ud}. The main benefit is that during learning \emph{additional information can be used} (e.g., global state, action, or rewards) and during execution this information is removed.

\item \textit{Parameter sharing.} 
Another frequent component in many \MDRL works is the idea of sharing parameters, i.e., training a single network in which agents share their weights. Note that, since agents could receive different observations (e.g., in partially observable scenarios), they can still behave differently. This method was proposed concurrently in different works~\cite{Gupta:2017to,Foerster:2016ud} and later it has been successfully applied in many others~\cite{Sukhbaatar:2016tk,Peng:vd,Foerster:2017ti,Sunehag:2018vua,Rashid:2018wy}.

\item \textit{Recurrent networks.} Recurrent neural networks (RNNs) enhanced neural networks with a memory capability, however, they suffer from the vanishing gradient problem, which renders them inefficient for long-term dependencies~\cite{pascanu2013difficulty}. However, RNN variants such as LSTMs~\cite{hochreiter1997long,Greff:2017gl} and GRUs (Gated Recurrent Unit)~\cite{chung2014empirical} addressed this challenge. In single-agent DRL, DRQN~\cite{Hausknecht:tq} initially proposed idea of using recurrent networks in single-agent \emph{partially observable} environments. Then, Feudal Networks~\cite{Vezhnevets:2017wl} proposed a hierarchical approach~\cite{dayan1993feudal}, \emph{multiple LSTM networks with different time-scales}, i.e., the observation input schedule is different for each LSTM network, to create a temporal hierarchy so that it can better address the long-term credit assignment challenge for RL problems. Recently, the use of recurrent networks has been extended to \MDRL to address the challenge of partially observability~\cite{Bansal:2017wl,Foerster:2016ud,Peng:vd,Omidshafiei:2017uk,Sunehag:2018vua,Rashid:2018wy,Raileanu:2018tb,Hong:2018vp,Rabinowitz:2018uf} for example, in FTW~\cite{Jaderberg:2018wx}, depicted in Figure~\ref{fig:ftw} and DRPIRQN~\cite{Hong:2018vp} depicted in Figure~\ref{fig:DPIRQN}. See Section~\ref{sec:challengesmdrl} for practical challenges (e.g., training issues) of recurrent networks in \MDRL.   

\item \textit{Overfitting in MAL.} 
In single-agent RL, agents can overfit to the environment~\cite{whiteson2011protecting}. A similar problem can occur in multiagent settings~\cite{johanson2011accelerating}, agents can overfit, i.e., an agent's policy can
easily get stuck in a local optima and the learned policy may be only locally optimal to other agents' current policies~\cite{li2019robust}. This has the effect of limiting the generalization of the learned policies~\cite{Lanctot:vn}. To reduce this problem, a solution is to have a set of policies (an ensemble) and learn from them or best respond to the mixture of them~\cite{Lanctot:vn,Lowe:2017wb,He:2016up}. Another solution has been to robustify algorithms --- a robust policy should be able to behave well even with strategies different from its training (better generalization)~\cite{li2019robust}.

\end{itemize}

\subsection{Benchmarks for MDRL}
\label{sec:benchmarksmdrl}

Standardized environments such as the Arcade Learning Environment (ALE)~\cite{bellemare2013arcade,machado2018revisiting} and OpenAI Gym~\cite{brockman2016openai} have allowed single-agent RL to move beyond toy domains. For DRL there are open-source frameworks that provide compact and reliable implementations of some state-of-the-art DRL algorithms~\cite{castro18dopamine}. Even though MDRL is a recent area, there are now a number of open sourced simulators and benchmarks to use with different characteristics, which we describe below.

\begin{figure}
\centering
\begin{subfigure}[]{.49\linewidth}
\centering
\includegraphics[scale=0.53]{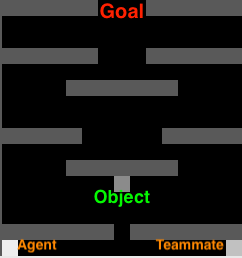}
\caption{Multiagent object transportation}
\label{fig:CMOTP}
\end{subfigure} 
\begin{subfigure}[]{.49\linewidth}
\centering
\includegraphics[scale=0.11]{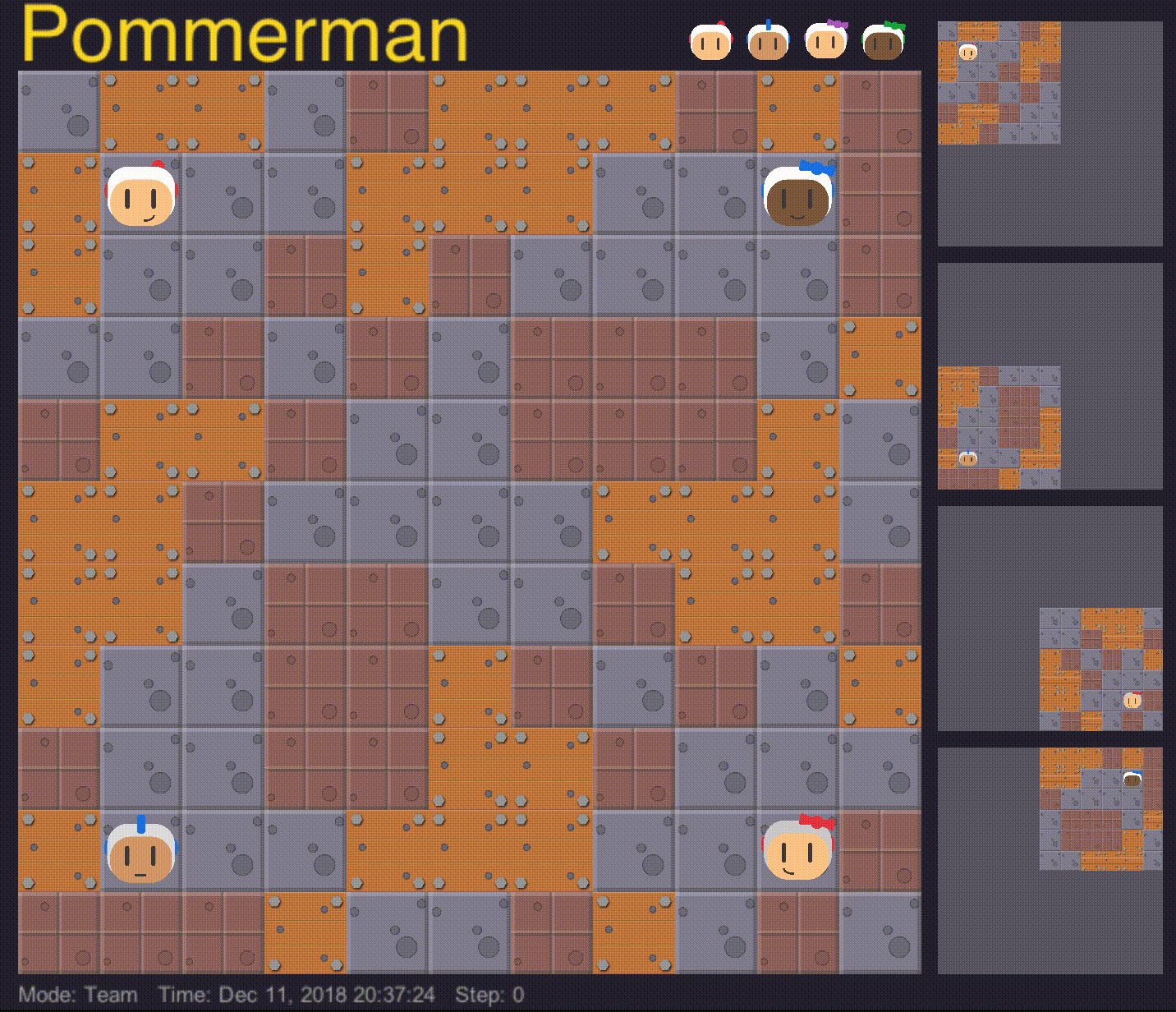}
\caption{Pommerman}
\label{fig:pommerman}
\end{subfigure}
\caption{(a) A fully cooperative benchmark with two agents, Multiagent Object Trasportation. (b) A mixed cooperative-competitive domain with four agents, Pommerman. For more \MDRL benchmarks see Section~\ref{sec:benchmarksmdrl}.}
\label{fig:benchmarks}
\end{figure}

\begin{itemize}
\item Fully Cooperative Multiagent Object Transporation Problems (CMOTPs)\footnote{\url{https://github.com/gjp1203/nui_in_madrl}} were originally presented by Busoniu et al.~\cite{Busoniu:2010ft} as a simple two-agent coordination problem in MAL. Palmer et al.~\cite{Palmer:2018wv} proposed  two pixel-based extensions to the original setting which include narrow passages that test the agents' ability to master fully-cooperative sub-tasks, stochastic rewards and noisy observations, see Figure~\ref{fig:CMOTP}. 

\item The Apprentice Firemen Game\footnote{\url{https://github.com/gjp1203/nui_in_madrl}} (inspired by the classic climb game~\cite{Claus:1998tb}) is another two-agent pixel-based environment that simultaneously confronts
learners with four pathologies in MAL: relative overgeneralization, stochasticity, the moving target problem, and alter exploration problem~\cite{palmer2018negative}.

\item Pommerman~\cite{pommerman} is a multiagent benchmark useful for testing cooperative, competitive and mixed (cooperative and competitive) scenarios. It supports partial observability and communication among agents, see Figure~\ref{fig:pommerman}. Pommerman is a very challenging domain from the exploration perspective as the rewards are very sparse and delayed~\cite{gao2019exploration}. A recent competition was held during NeurIPS-2018\footnote{\url{https://www.pommerman.com/}} and the top agents from that competition are available for training purposes.

\item Starcraft Multiagent Challenge~\cite{samvelyan2019starcraft} is based on the real-time strategy
game StarCraft II and focuses on micromanagement challenges,\footnote{\url{https://github.com/oxwhirl/smac}} that is, fine-grained
control of individual units, where each unit is controlled by an independent agent that must act based on local observations. It is accompanied by a \MDRL framework including state-of-the-art algorithms (e.g., QMIX and COMA).\footnote{\url{https://github.com/oxwhirl/pymarl}}

\item The Multi-Agent Reinforcement Learning
in Malm\"o (MARL\"O) competition~\cite{perez2019multi}  is another multiagent challenge with multiple cooperative 3D games\footnote{\url{https://github.com/crowdAI/marlo-single-agent-starter-kit/}} within Minecraft. The scenarios were created with the open source Malm\"o platform~\cite{johnson2016malmo}, providing examples of how a wider range of multiagent cooperative, competitive and mixed scenarios can be experimented on within Minecraft.

\item Hanabi is a cooperative multiplayer card game (two to five players). The main characteristic of the game is that players do not observe their own cards but other players can reveal information about them. This makes an interesting challenge for learning algorithms in particular in the context of self-play learning and ad-hoc teams~\cite{Stone:2010wi,bowling2005coordination,Albrecht:2013tea}. The Hanabi Learning Environment~\cite{bard2019hanabi} was recently released\footnote{\url{https://github.com/deepmind/hanabi-learning-environment}} and it is accompanied with a baseline (deep RL) agent~\cite{hessel2018rainbow}.

\item Arena~\cite{arena2019} is platform for multiagent research\footnote{\url{https://github.com/YuhangSong/Arena-BuildingToolkit}} based on the Unity engine~\cite{juliani2018unity}. It has 35 multiagent games (e.g., social dilemmas) and supports communication among agents. It has basseline implementations of recent DRL algorithms such as independent PPO learners.

\item MuJoCo Multiagent Soccer~\cite{liu2019emergent} uses the MuJoCo physics engine~\cite{Todorov:2012vi}. The environment simulates a 2 vs.\ 2 soccer game with agents having a 3-dimensional action space.\footnote{\url{https://github.com/deepmind/dm_control/tree/master/dm_control/locomotion/soccer}} 

\item Neural MMO~\cite{suarez2019neural} is a research platform\footnote{\url{https://github.com/openai/neural-mmo}} inspired by the human game genre of Massively Multiplayer Online (MMO) Role-Playing Games. These games involve a large, variable number of players competing to survive.
\end{itemize}

\subsection{Practical challenges in MDRL}
\label{sec:challengesmdrl}

In this section we take a more critical view with respect to \MDRL and highlight different practical challenges that already happen in DRL and that are likely to occur in MDRL such as reproducibility, hyperparameter tuning, the need of computational resources and conflation of results. We provide pointers on how we think those challenges could be (partially) addressed.

\paragraph{Reproducibility, troubling trends and negative results}

Reproducibility is a challenge in RL which is only aggravated in DRL due to different sources of stochasticity: baselines, hyperparameters, architectures~\cite{Henderson:2018tg,Nagarajan:2018} and random seeds~\cite{clary2019let}. Moreover, DRL does not have common practices for statistical testing~\cite{forde2019scientific} which has led to bad practices such as only reporting the results when algorithms perform well, sometimes referred as \emph{cherry picking}~\cite{azziz2019} (Azizzadenesheli also describes \emph{cherry planting} as adapting an environment to a specific algorithm~\cite{azziz2019}). We believe that together with following the advice on how to design experiments and report results~\cite{lipton2018troubling}, the community would also benefit from reporting \emph{negative results}~\cite{rosenthal1979file,sculley2018winner,forde2019scientific,gencoglu2019hark} for carefully designed hypothesis and experiments.\footnote{This idea was initially inspired by the Workshop ``Critiquing and Correcting Trends in Machine Learning" at NeurIPS 2018 where it was possible to submit \emph{Negative results} papers: ``Papers which show failure modes of existing algorithms or suggest new approaches which one might expect to perform well but which do not. The aim is to provide a venue for work which might otherwise go unpublished but which is still of interest to the community." \url{https://ml-critique-correct.github.io/}} However, we found very few papers with this characteristic\cite{azizzadenesheli2018surprising,lyle2019comparative,kartal2018using} --- we \revise{note that this is not encouraged in the ML community; moreover, negative results reduce the chance of paper acceptance~\cite{lipton2018troubling}. In this regard, we ask the community to reflect on these practices and find ways to remove these obstacles.}

\paragraph{Implementation challenges and hyperparameter tuning}
One problem is that canonical implementations of DRL algorithms often contain additional non-trivial optimizations --- these are sometimes necessary for the algorithms to achieve good performance~\cite{ilyas2019truly}. A recent study by Tucker et al.~\cite{tucker2018mirage} found that several published works on action-dependant baselines contained bugs and errors --- those were the real reason of the high performance in the experimental results, not the proposed method.
Melis et al.~\cite{melis2017state} compared a series of works with increasing innovations in network architectures and the vanilla LSTMs~\cite{hochreiter1997long} (originally proposed in 1997). The results showed that, when properly tuned, LSTMs outperformed the more recent models. In this context, Lipton and Steinhardt noted that the community may have benefited more by learning the details of the hyperparameter tuning~\cite{lipton2018troubling}. A partial reason for this surprising result might be that this type of networks are known for being difficult to train~\cite{pascanu2013difficulty} and there are recent works in DRL that report problems when using recurrent networks~\cite{10.1007/978-3-319-71682-4_5,deeprlpongfrompixels,firoiu2017beating,gao2019skynet}. Another known complication is catastrophic forgetting (see Section~\ref{sec:deeprl}) with recent examples in DRL~\cite{Raghu:2018tg,van2016deep} --- we expect that these issues would likely occur in \MDRL. The effects of hyperparameter tuning were analyzed in more detail in DRL by Henderson et al.~\cite{Henderson:2018tg}, \revise{who arrived} at the conclusion that hyperparameters can have significantly different effects across algorithms (they tested TRPO, DDPG, PPO and ACKTR) and environments since there is an intricate interplay among them~\cite{Henderson:2018tg}. The authors urge the community to report \emph{all} parameters used in the experimental evaluations for accurate comparison --- we encourage a similar behavior for \MDRL.
\revise{Note that hyperparameter tuning is related to the troubling trend of cherry picking in that it can show a carefully picked set of parameters that  make an algorithm work (see previous challenge). Lastly, note that hyperparameter tuning is computationally very expensive, which relates to the connection with the following challenge of computational demands.}

\paragraph{Computational resources}

Deep RL usually requires millions of interactions for an agent to learn~\cite{amodei2ai}, i.e., low sample efficiency~\cite{yu2018towards}, which highlights the need for large computational infrastructure in general. The original A3C implementation~\cite{mnih2016asynchronous} uses 16 CPU workers for 4 days to learn to play an Atari game with a total of 200M training frames\footnote{It is sometimes unclear in the literature what is the meaning of frame due to the ``frame skip" technique. It is therefore suggested to refer to ``game frames" and ``training frames"~\cite{such2017deep}.} (results are reported for 57 Atari games). Distributed PPO used 64 workers (presumably one CPU per worker, although this is not clearly stated in the paper) for 100 hours (more than 4 days) to learn locomotion tasks~\cite{Heess:2017ts}. 
In \MDRL, for example, the Atari Pong game, agents were trained for 50 epochs, 250k time steps each, for a total of 1.25M training frames~\cite{Tampuu:2017fc}. The FTW agent~\cite{Jaderberg:2018wx} uses 30 agents (processes) in parallel and every training game lasts for five minues; FTW agents were trained for approximately 450K games $\approx$4.2 years. 
These examples highlight the computational demands sometimes needed within DRL and \MDRL. 

Recent works have reduced the learning of an Atari game to minutes \revise{(Stooke and Abbeel~\cite{stooke2018accelerated} trained DRL agents in less than one hour with hardware consisting of 8 GPUs and 40 cores).} However, this is (for now) the exception and computational infrastructure is a major bottleneck for doing DRL and MDRL, especially for those who do not have such large compute power (e.g., \revise{most companies and most academic research groups})~\cite{beeching2019deep,sculley2018winner}.\footnote{One recent effort by Beeching et al.~\cite{beeching2019deep} proposes to use only ``mid-range hardware" (8 CPUs and 1 GPU) to train deep RL agents.} Within this context we propose two ways to address this problem. (1) Raising awareness: For DRL we found few works that study the computational demands of recent algorithms~\cite{babaeizadeh2016reinforcement,amodei2ai}. For \MDRL most published works do not provide information regarding computational resources used such as CPU/GPU usage, memory demands, and wall-clock computation. Therefore, the first way to tackle this issue is by raising awareness and encouraging authors to report metrics about computational demands for accurately comparison and evaluation. (2) Delve into algorithmic contributions. Another way to address these issues is to prioritize the algorithmic contribution for the new MDRL algorithms rather than the computational resources spent. Indeed, for this to work, it needs to be accompanied with high-quality reviewers.

We have argued to raise awareness on the computational demands and report results, however, there is still the open question on \emph{how} and \emph{what} to measure/report. 
There are several dimensions to measure efficiency: sample efficiency is commonly measured by counting state-action pairs used for training; computational efficiency could be measured by number of CPUs/GPUs and days used for training. How do we measure the impact of other resources, \revise{such as} external data sources or annotations?\footnote{NeurIPS 2019 hosts the ``MineRL Competition on Sample Efficient Reinforcement Learning using Human Priors" where the primary goal of the competition is to foster the development of algorithms which can efficiently leverage human demonstrations to drastically reduce the number of samples needed to solve complex, hierarchical, and sparse environments~\cite{guss2019minerl}.} Similarly, do we need to differentiate the computational needs of the algorithm itself versus the environment it is run in? We do not have the answers, however, we point out that current standard metrics might not be entirely comprehensive.
     
In the end, we believe that high compute based methods act as a frontier to showcase benchmarks~\cite{alphastarblog,openfive}, i.e., they show what results are possible as data and compute is scaled up (e.g., OpenAI Five generates 180 years of gameplay data each day using 128,000 CPU cores and 256 GPUs~\cite{openfive}; AlphaStar uses 200 years of Starcraft II gameplay~\cite{alphastarblog}); however, lighter compute based algorithmic methods can also yield significant contributions to better tackle real-world problems.
    
\paragraph{Occam's razor and ablative analysis}
Finding the simplest context that exposes the innovative research idea remains challenging, and if ignored leads to a conflation of fundamental research (working principles in the most abstract setting) and applied research (working systems as complete as possible). In particular, some deep learning papers are presented as learning from pixels without further explanation, while object-level representations would have already exposed the algorithmic contribution. This still makes sense to remain comparable with established benchmarks (e.g., OpenAI Gym~\cite{brockman2016openai}), but less so if custom simulations are written without open source access, as it introduces unnecessary variance in pixel-level representations and artificially inflates computational resources (see previous point about \emph{computational resources}).\footnote{Cuccu, Togelius and Cudr\'e-Mauroux achieved state-of-the-art policy learning in Atari games with only 6 to 18 neurons~\cite{cuccu2019playing}. The main idea was to decouple image processing from decision-making.} %
In this context there are some notable exceptions where the algorithmic contribution is presented in a minimal setting and then results are scaled into complex settings: LOLA~\cite{Foerster:2018to} first presented a minimalist setting with a two-player two-action game and then with a more complex variant; similarly, QMIX~\cite{Rashid:2018wy} presented its results in a two-step (matrix) game and then in the more involved Starcraft II micromanagement domain~\cite{samvelyan2019starcraft}.

\subsection{Open questions}
\label{sec:bridging:questions}

Finally, here we present some open questions for MDRL and point to suggestions on how to approach them. We believe that there are solid ideas in earlier literature and we refer the reader to Section~\ref{sec:bridging:examples} to avoid deep learning amnesia.

\begin{itemize}

\item On the challenge of sparse and delayed rewards. 

Recent \MDRL competitions and environments have complex scenarios where many actions are taken before a reward signal is available (see Section~\ref{sec:benchmarksmdrl}). This sparseness is already a challenge for RL~\cite{sutton2018reinforcement,ecoffet2019go} where approaches such as count-based exploration/intrinsic motivation~\cite{bellemare2016unifying,brafman2002r,strehl2008analysis,schmidhuber1991possibility, barto2013intrinsic} and hierarchical learning~\cite{kulkarni2016hierarchical,Dietterich:2000gl,Schaul:2015ty} have been proposed to address it --- in \MDRL this is even more problematic since the agents not only need to learn basic behaviors (like in DRL), but also to learn the strategic element (e.g., competitive/collaborative) embedded in the multiagent setting. To address this issue, recent \MDRL approaches applied \emph{dense} rewards~\cite{Ng:1999vl,konidaris2006autonomous,mahadevan1992automatic} (a concept originated in RL) at each step to allow the agents to learn basic motor skills and then decrease these \emph{dense} rewards over time in favor of the environmental reward~\cite{Bansal:2017wl}, see Section~\ref{sec:emergentBehaviors}. Recent works like OpenAI Five~\cite{openfive} uses hand-crafted intermediate rewards to accelerate the learning and FTW~\cite{Jaderberg:2018wx} lets the agents learn their internal rewards by a hierarchical two-tier optimization. In \textit{single agent} domains, RUDDER~\cite{arjona2018rudder} has been recently proposed for such delayed sparse reward problems. RUDDER generates a new MDP with \emph{more intermediate rewards} whose optimal solution is still an optimal solution to the original MDP. This is achieved by using LSTM networks to redistribute the original sparse reward to earlier state-action pairs and automatically provide reward shaping. %
How to best extend RUDDER to multiagent domains is an open avenue of research.

\item On the role of self-play. %

Self-play is a cornerstone in MAL with impressive results~\cite{Bowling:2002vva,Hu:2003wb,Bowling:2005vi,Conitzer:2006du,Greenwald:2003uy}. While notable results had also been shown in \MDRL~\cite{Heinrich:69lSVXjW,Bowling:2015ub}, recent works have also shown that \emph{plain} self-play does not yield the best results. However, adding diversity, i.e., evolutionary methods~\cite{back1996evolutionary,de2006evolutionary,rosin1997new,lehman2008exploiting} or sampling-based methods, have shown good results~\cite{Bansal:2017wl,Jaderberg:2018wx,leibo2018malthusian}. A drawback of these solutions is the additional computational requirements since they need either parallel training (more CPU computation) or memory requirements. Then, it is still an open problem to improve the computational efficiency of these previously proposed successful methods, i.e., achieving similar training stability with smaller population sizes that uses fewer CPU workers in MAL and \MDRL (see Section~\ref{sec:challengesmdrl} and Albrecht et al.~\cite[Section 5.5]{Albrecht:2018dp}).

\item On the challenge of the combinatorial nature of \MDRL.

Monte Carlo tree search (MCTS)~\cite{browne2012survey} has been the backbone of \revise{the} major breakthroughs \revise{behind} AlphaGo~\cite{Silver:2016hl} and AlphaGo Zero~\cite{silver2017mastering} that combined search and DRL. A recent work~\cite{vodopivec2017monte} has outlined how search and RL can be better combined for potentially new methods. However, for multiagent scenarios, there is an additional challenge of the exponential growth of all the agents' action spaces for centralized methods~\cite{kartal2015stochastic}. One way to tackle this challenge within multiagent scenarios is the use of search parallelization~\cite{kartal2016monte,best2019dec}. Given more scalable planners, there is room for research in combining these techniques in \MDRL settings.

To learn complex multiagent interactions some type of abstraction~\cite{DeHauwere:2010vq} is often needed, for example, factored value functions~\cite{guestrin2002coordinated,guestrin2003efficient,guestrin2002multiagent,kok2004sparse,amato2015scalable,Oliehoek:2018tn} (see QMIX and VDN in Section~\ref{sec:learningCooperation} for recent work in MDRL) try to exploit independence among agents through (factored) structure; however, in MDRL there are still open questions such as understanding their representational power~\cite{castellini2019representational} (e.g., the accuracy of the learned Q-function approximations) and how to learn those factorizations, where ideas from transfer planning techniques could be useful~\cite{oliehoek2013approximate,van2016coordinated}. In transfer planning the idea is to define a simpler ``source problem" (e.g., with fewer agents), in which the agent(s) can plan~\cite{oliehoek2013approximate} or learn~\cite{van2016coordinated}; since it is less complex than the real multiagent problem, issues such as the non-stationarity of the environment can be reduced/removed. Lastly, another related idea are \emph{influence} abstractions~\cite{becker2004solving,oliehoek2012influence,HernandezLeal:2017wd}, where instead of learning a complex multiagent model, these methods try to build smaller models based on the influence agents can exert on one another. While this has not been sufficiently explored in actual multiagent settings, there is some evidence that these ideas can lead to effective inductive biases, improving effectiveness of DRL in such local abstractions~\cite{Suau19ALA}.

\end{itemize}

\section{Conclusions}
\label{sec:conclusions}

Deep reinforcement learning has shown recent success on many fronts~\cite{Mnih:2015jp,Silver:2016hl,Moravcik:2017jb} and a natural next step is to test multiagent scenarios. However, learning in multiagent environments is fundamentally more difficult due to non-stationarity, the increase of dimensionality, and the credit-assignment problem, among other factors~\cite{Stone:2000ww,Busoniu:2008bo,HernandezLeal:2017wd,Bowling:2002vva,Tumer:2007vs,Wei:2018wu,palmer2018negative}.

This survey provides broad overview of recent works in the emerging area of Multiagent Deep Reinforcement Learning (\MDRL). First, we categorized recent works into four different topics: emergent behaviors, learning communication, learning cooperation, and agents modeling agents. Then, we exemplified how key components (e.g., experience replay and difference rewards) originated in RL and MAL need to be adapted to work in MDRL. We provided general lessons learned applicable to \MDRL, pointed to recent multiagent benchmarks and highlighted some open research problems. Finally, we also reflected on the practical challenges such as computational demands and reproducibility in \MDRL.

\revise{Our conclusions of this work are that while} the number of works in DRL and \MDRL are notable and represent important milestones for AI, at the same time we acknowledge there are also open questions in both (deep) single-agent learning~\cite{Darwiche:2018,machado2018revisiting,torrado2018deep,ilyas2019truly} and multiagent learning~\cite{ortega2018modeling,pmlr-v80-yang18d,Grover:2018wl,ling2018game,Omidshafiei2019neuralrd,khada2019merl}. Our view is that there are practical issues within \MDRL that hinder its scientific progress: the necessity of high compute power, complicated reproducibility (e.g., hyperparameter tuning), and the lack of sufficient encouragement for publishing negative results. However, we remain highly optimistic of the multiagent community and hope this work serves to raise those issues, encounter good solutions, and ultimately take advantage of the existing literature and resources available to move the area in the right direction.

\section*{Acknowledgements}
We would like to thank Chao Gao, Nidhi Hegde, Gregory Palmer, Felipe Leno Da Silva and Craig Sherstan for reading earlier versions of this work and providing feedback, to April Cooper for her visual designs for the figures in the article, to Frans Oliehoek, Sam Devlin, Marc Lanctot, Nolan Bard, Roberta Raileanu, Angeliki Lazaridou, and Yuhang Song for clarifications in their areas of expertise, to Baoxiang Wang for his suggestions on recent deep RL works, to Michael Kaisers, Daan Bloembergen, and Katja Hofmann for their comments about the practical challenges of MDRL, \revise{and} to the editor and three anonymous reviewers whose comments and suggestions increased the quality of this work.

\bibliographystyle{elsarticle-num}   
\bibliography{bibAll,bibExtra}
\end{document}